\documentclass[11pt]{article}
\pdfoutput=1
\usepackage{jheppub}
\usepackage{graphicx}
\usepackage{amssymb}
\usepackage{amsmath}
\usepackage{amsfonts}
\usepackage{latexsym}
\usepackage{hyperref}
\usepackage{physics}
\usepackage{bbm}
\usepackage{empheq}
\usepackage{tikz}
\usepackage{cancel}
\usepackage{hhline}
\usepackage{caption}
\usepackage{soul}
\usepackage{float}
\usepackage{multirow}
\usepackage{hhline}
\usepackage{algorithm}
\usepackage{algpseudocode}
\usepackage{appendix}
\usepackage{subcaption}

\usepackage[T1]{fontenc}

\usetikzlibrary{arrows.meta}
\usetikzlibrary{decorations.markings}
\tikzset{->-/.style={decoration={
			markings,
			mark=at position #1 with {\arrow{>}}},postaction={decorate}}}

\def\ri{{\rm i}}

\newcommand{\R}{\mathbb{R}}

\def\rd{{\rm d}}

\def\N{\ensuremath{\mathcal{N}}}
\def\O{\ensuremath{\mathcal{O}}}
\def\I{\ensuremath{\mathcal{I}}}

\def\etah{\ensuremath{\tfrac{\eta}{2}}}
\def\Veta{\ensuremath{\mathcal{V}_\eta}}

\makeatletter
\newcommand*{\rom}[1]{\expandafter\@slowromancap\romannumeral #1@}
\makeatother

\begin{document}

% Front page here
\thispagestyle{empty}

~\\[-1.5cm]

\begin{center}
\vskip 1.2truecm 
{\Large\bf
%\titleline
%
{\LARGE Correlation functions for non-conformal \\\vskip 0.3truecm  D$p$-brane holography}
}\\
\vskip 2.25truecm
	{\bf Nikolay Bobev, Guillermo Mera \'Alvarez, Hynek Paul \\
	}
		\vskip 0.4truecm
 
	{\it	Instituut voor Theoretische Fysica, KU Leuven, \\
	Celestijnenlaan 200D, B-3001 Leuven, Belgium\\\vskip .2truecm
		\vskip .2truecm   }
	{\it	Leuven Gravity Institute, KU Leuven, \\
	Celestijnenlaan 200D box 2415, 3001 Leuven, Belgium\\\vskip .2truecm
		\vskip .2truecm   }

	\vskip .2truecm

\href{nikolay.bobev@kuleuven.be}{{\tt nikolay.bobev}}\,,\,\href{mailto:guillermo.meraalvarez@kuleuven.be}{{\tt guillermo.meraalvarez}}\,,\,\href{mailto:hynek.paul@kuleuven.be}{{\tt hynek.paul@kuleuven.be}}

\end{center}

\vskip 1.25truecm

%%%%%%%%%%%%%%%%%%%%%%%%%%%%
\centerline{\bf Abstract}
\vskip .4truecm

\noindent We use holography to study correlation functions of local operators in maximally supersymmetric Yang-Mills theories arising on the world-volume of D$p$-branes in the large-$N$ and strong-coupling limit. The relevant supergravity backgrounds obtained from the near-horizon limit of the D$p$-branes enjoy a scaling similarity, which leads to an auxiliary AdS space of fractional dimension. This suggests that holographic correlation functions in this setup can be computed by integrating standard CFT correlators over the auxiliary extra dimensions. We apply this prescription to analytically compute two- and three-point correlators of scalar operators. The resulting two-point functions take a familiar CFT form but with shifted conformal dimensions, while the three-point correlators have a much more involved position dependence which we calculate explicitly in terms of a sum of Appell functions.

%%%%%%%%%%%%%%%%%%%%%%%%%%%%

\newpage
\setcounter{page}{1}\setcounter{footnote}{0}
\setcounter{tocdepth}{2}
\tableofcontents

%%%%%%%%%%%%%
\section{Introduction}
\label{sec:intro}
%%%%%%%%%%%%%

The AdS/CFT correspondence is an invaluable tool for the explicit calculation of correlation functions in strongly coupled CFTs, see \cite{DHoker:2002nbb} for a review and references to the original literature. It is clearly important to extend this powerful duality and find holographic methods to calculate correlators in non-conformal strongly interacting QFTs. Since the landscape of non-conformal QFTs is unwieldy, it may be too ambitious to look for such a generalization without further guidance. As is often the case, string theory and its low-energy supergravity limit suggest a natural set of non-conformal theories and point to their holographic realization.

Indeed, soon after the seminal work \cite{Maldacena:1997re} that established the duality between the~AdS$_5\times S^5$ near-horizon limit of the supergravity solution describing $N$ coincident D3-branes and the 4d ${\rm SU}(N)$ $\mathcal{N}=4$ SYM theory, it was proposed that the holographic duality should also hold for D$p$-branes with $p\neq 3$ \cite{Itzhaki:1998dd}. On one hand, the low-energy dynamics of the open strings ending on $N$ coincident D$p$-branes\footnote{We will take $p\leq 6$ in this paper and will often assume that $p\neq 3$ to exclude the special case of D3-branes and the corresponding $\mathcal{N}=4$ SYM CFT.} is described by the maximally supersymmetric ${\rm U}(N)$ Yang-Mills (YM) theory in $d=p+1$ dimensions. On the other hand, when $N$ is large the backreaction of the D$p$-branes cannot be ignored and there is an effective closed string description of the system in terms of an explicit solution of IIA/B supergravity. It was then proposed in~\cite{Itzhaki:1998dd} that, after taking a near-horizon (or decoupling) limit, the supergravity background is holographically dual to the strong-coupling limit of the planar ${\rm SU}(N)$ Yang-Mills theory.\footnote{As usual, in the near-horizon limit the center of mass motion of the stack of $N$ D$p$-branes decouples from the dynamics. This is reflected by the change of the gauge group from ${\rm U}(N)$ to ${\rm SU}(N)$.}

Despite this clear route to extending the holographic correspondence, the non-conformal D$p$-brane holographic duality has not enjoyed as much quantitative success as its AdS/CFT counterpart. On the gravity side, the reduced symmetry and the lack of a mathematically well-understood asymptotic structure at spatial infinity, akin to the Fefferman-Graham expansion, complicates the analysis of holographic correlation functions and the construction of non-trivial backgrounds, like holographic RG flows and black holes. On the field theory side, the lack of conformal invariance leads to reduced calculational control of the QFT observables, despite the large amount of supersymmetry. In addition, our incomplete understanding of string theory on backgrounds with RR fluxes further limits the number of available techniques that can shed light on this problem. Nevertheless, progress in top-down non-conformal D$p$-brane holography has been achieved on several fronts. This includes the development of a holographic renormalization method in \cite{Kanitscheider:2008kd,Kanitscheider:2009as,Wiseman:2008qa} which paves the way for the calculation of holographic correlation functions and supergravity on-shell actions. Moreover, the advances in supersymmetric localization techniques, see \cite{Pestun:2016zxk} for a review, led to the calculation of the maximally supersymmetric YM path integral on $S^{p+1}$ in terms of a matrix model, see \cite{Blau:2000xg,Minahan:2015jta}. This result in turn leads to precision tests of non-conformal holography in which the supergravity and probe string on-shell actions in the spherical D$p$-brane supergravity solutions of \cite{Bobev:2018ugk} were shown to agree with the free energy and the supersymmetric circular Wilson line expectation value on $S^{p+1}$ \cite{Bobev:2019bvq,Astesiano:2024sgi,Bobev:2024gqg}. The special case of D0-branes allows for additional tests of the holographic correspondence in a thermal setting, since the BFSS matrix quantum mechanics and its BMN generalization \cite{Banks:1996vh,Berenstein:2002jq} can be modelled on the lattice at finite temperature. These lattice results can then be compared with the dual D0-brane black hole, see \cite{Hanada:2008ez,Catterall:2007fp,Catterall:2008yz,Hanada:2011fq,Hanada:2013rga,Costa:2014wya,Biggs:2023sqw} for an incomplete selection of relevant work in this direction. Generalizations of this lattice approach for D1- and D2-branes have also been explored, see \cite{Catterall:2010fx,Catterall:2017lub,Catterall:2020nmn}.

The main goal of this work is to continue the exploration of precision top-down D$p$-brane holography by focusing on the holographic calculation of correlation functions. At first sight, the problem appears conceptually straightforward. One needs to calculate the various bulk/boundary-to-bulk/boundary propagators in the D$p$-brane supergravity solutions of \cite{Itzhaki:1998dd} and then proceed to evaluate the corresponding Witten diagrams. This blueprint appears to be deceptively simple and the analysis has not been performed even for the simplest non-trivial case of 3pt correlation functions of scalar operators. Here we aim to address this question and compute the explicit form of the 3pt-functions for all D$p$-branes in the supergravity limit.

To make progress we take advantage of two related structural features of the near-horizon D$p$-brane supergravity solutions of \cite{Itzhaki:1998dd}. As emphasized in \cite{Boonstra:1998mp,Jevicki:1998ub,Sekino:1999av,Smilga:2008bt,Wiseman:2013cda,Biggs:2023sqw}, there is a scaling similarity of the classical supergravity action which constrains the form of the on-shell action and the behavior of linearized fluctuations around a fixed supergravity solution. Moreover, it was shown in \cite{Boonstra:1998mp,Jevicki:1998ub,Sekino:1999av,Kanitscheider:2008kd,Kanitscheider:2009as,Biggs:2023sqw} that the D$p$-brane supergravity solutions of \cite{Itzhaki:1998dd} enjoy an auxiliary AdS structure. One way to see this is to note that the metric of the supergravity D$p$-brane solutions has the warped product form $w({\rm AdS}_{p+2} \times S^{8-p})$ where the warp factor is related to the running dilaton of the solution and depends on the AdS radial coordinate. Alternatively, one can view the 10d supergravity background as arising from a dimensional reduction of an auxiliary higher-dimensional AdS$_{2+p+\eta}\times S^{8-p}$ background where $\eta = \frac{(3-p)^2}{5-p}$ is in general a fractional number. Importantly, this uplifted solution in $10+\eta$ dimensions has a vanilla AdS factor without any warping. This higher-dimensional AdS structure suggests an indirect way to compute the D$p$-brane correlation functions. One can take a CFT correlation function in $p+1+\eta$ dimensions and then integrate it over the $\mathbb{R}^{\eta}$ subspace of $\R^{p+1+\eta}$ to obtain the $(p+1)$-dimensional correlator of interest. Indeed, this approach was used in \cite{Kanitscheider:2009as,Biggs:2023sqw} for the calculation of 2pt-functions in the D$p$-branes supergravity solutions.\footnote{See \cite{Batra:2025ivy} for a recent application of the auxiliary conformal structure in the context of giant gravitons for D$p$-branes.} This is also the method we employ in the present work.

Guided by the auxiliary CFT$_{p+1+\eta}$ structure suggested by the bulk supergravity description, we simply postulate that the $(p+1)$-dimensional correlation functions of local operators in the maximally supersymmetric YM theory can be obtained by integrating local CFT$_{p+1+\eta}$ correlation functions over $\mathbb{R}^{\eta}$. Applying this procedure to 2pt-functions of scalar operators leads to 2pt-correlators on $\mathbb{R}^{p+1}$ that take the standard form of a CFT$_{p+1}$ 2pt-function albeit with shifted operator scaling dimensions. As usual in holography, the operator scaling dimension is determined by the mass of the dual linearized perturbation in the D$p$-brane supergravity solution. This result may be somewhat surprising since the underlying SYM theory is not conformally invariant for $p\neq 3$. Indeed, the shift in the operator scaling dimensions is controlled by the parameter $\eta$ and is ultimately responsible for the breaking of the conformal invariance in the 2pt-function for $p\neq 3$. This scaling behavior of the D$p$-brane 2pt-functions is closely related to the scaling similarity of the classical 10d supergravity action. It constitutes an important holographic prediction for the strongly coupled dynamics of SYM theory in the planar limit which is difficult to confirm with field theory methods.

The analysis of the 3pt-functions of scalar operators using this method is more challenging. We start with the standard form of the 3pt-correlation function of a CFT on $\mathbb{R}^{p+1+\eta}$ and then need to perform three integrals over the location of the three points in the $\mathbb{R}^{\eta}$ subspace. We can use translation invariance to reduce this triple integral to a non-trivial double integral. It turns out that integrals of this form have been studied in the literature on perturbative two-loop calculations in QFT~\cite{Davydychev:1992mt}. We borrow and adapt these results to our setup to find an explicit analytic form of the $(p+1)$-dimensional scalar 3pt-functions expressed as a sum of four Appell $F_4$ hypergeometric functions with arguments determined by the position $x_i$ of the three points on $\mathbb{R}^{p+1}$, the scaling dimensions $\Delta_i$ of the scalar operators and the parameter $\eta$. This is the main result of our work and constitutes a non-trivial prediction for the kinematic structure of 3pt-correlation functions of scalar operators in the maximally supersymmetric YM theory dual to the D$p$-brane background. In the absence of an alternative QFT or supergravity method to do this 3pt-function calculation and independently check our results, we resort to several internal consistency checks. We show explicitly that the complicated expression for the 3pt-function in terms of Appell $F_4$ functions reduces to the familiar CFT$_{p+1}$ 3pt-correlator in the $\eta \to 0$ limit. Furthermore, for any value of $\eta$, our result is invariant under the permutation of the three points, something far from obvious from the explicit Appell function form. Finally, for extremal arrangements of the operator dimensions, $\Delta_i=\Delta_j+\Delta_k$, or in the coincident point limit $x_i \to x_j$, we find that the Appell function form of the correlator reduces to a product of two-point functions, as expected on general grounds.

We continue in the next section with a summary of the well-known supergravity solutions describing the backreaction of coincident D$p$-branes and we discuss the scaling similarity and the corresponding auxiliary AdS structure associated with these backgrounds. We then proceed to utilize this auxiliary conformal structure in Section~\ref{sec:intCFT} where we derive our main results for the explicit form of the 2pt- and 3pt-functions and perform several consistency checks. We conclude in Section~\ref{sec:discussion} with a short discussion on some open questions and potential avenues for future work.

\medskip

\noindent\textit{Note added:} While we were preparing this manuscript, the paper~\cite{Bigss:2025} appeared on the arXiv. The authors of \cite{Bigss:2025} calculate 2pt- and 3pt-functions of the BFSS matrix quantum mechanics on the world-volume of D0-branes using the auxiliary AdS structure in the supergravity limit and the corresponding bulk Witten diagrams. The approach in~\cite{Bigss:2025} is somewhat complementary to the $p=0$ specialization of our analysis and it will be nice to perform a detailed comparison between the two sets of results. 

\medskip

\noindent\textit{Note added in v2:} Our result for the 2pt-functions for $p=0$ in~\eqref{eq:2ptreseta} is in agreement with the results in~\cite{Bigss:2025}. For $p=0$, the OPE limit of our general result for the 3pt-function presented in~\eqref{eq:OPE_limit} also appears to agree with the result for the “squeezed limit” of~\cite{Bigss:2025}.

%%%%%%%%%%%%%%
\section{D$p$-branes, scaling similarity, and auxiliary AdS}
\label{sec:auxAdS}
%%%%%%%%%%%%%%

The low-energy supergravity limit of type IIA/B string theory admits $p$-brane solutions that naturally source the $C_{p+1}$ RR fields \cite{Horowitz:1991cd}. These objects are characterized by their electric charge density $\mu_{p}$ and tension $T_{p}$. In the extremal limit $T_p=\mu_p$ these $p$-brane solutions preserve 16 of the 32 supercharges of the supergravity theory and can be interpreted as the backreaction of D$p$-branes. These are the supergravity solutions that form the basis of the holographic duality we want to study~\cite{Itzhaki:1998dd}.

The type IIA/B supergravity background of interest is\footnote{We follow the conventions used in \cite{Bobev:2018ugk} and present the solution in string frame.}
\begin{equation}\label{eq:10dsolR10}
\begin{split}
ds^2_{10} &= H^{-1/2} ds^2_{p+1} + H^{1/2}(dr^2 + r^2 d\Omega^2_{8-p})\,,\\
e^{\Phi}&= g_s H^{\frac{(3-p)}{4}}\,, \quad\qquad C_{p+1} = (g_sH)^{-1} {\rm vol}_{p+1}\,.
\end{split}
\end{equation}
Here $g_s$ is the string coupling parameter, $\Phi$ is the dilaton, $C_{p+1}$ is the RR potential, $ds^2_{p+1}$ is the metric on $\mathbb{R}^{1,p}$, and $d\Omega^2_{8-p}$ is the Einstein metric on $S^{8-p}$. The solution preserves 16 supercharges for any harmonic function $H$ on the $\mathbb{R}^{9-p}$ spanned by $r$ and $d\Omega^2_{8-p}$. The choice of $H$ determines the distribution of the D$p$-branes and for $N$ coincident branes it is given by
\begin{equation}\label{eq:Hcoincid}
H = 1 + \frac{g_s N}{\mu_{6-p}V_{6-p}r^{7-p}}\,,
\end{equation}
where we have defined
\begin{equation}\label{eq:muVdef}
\mu_{n} = (2\pi)^{-p} \ell_s^{-p-1} \,,  \qquad\qquad V_n =\frac{2 \pi^{\frac{n+1}{2}}}{\Gamma(\frac{n+1}{2})}\,, 
\end{equation}
and $\ell_s$ is the string length. The coupling constant of the dual $(p+1)$-dimensional SYM theory is expressed in terms of the string theory parameters as
\begin{equation}\label{eq:gYMdef}
g^2_{\rm YM} = \frac{2\pi g_s}{(2\pi \ell_s)^{3-p} }\,.
\end{equation}
As expected from the physics of the dual ${\rm SU}(N)$ SYM theory, $g^2_{\rm YM}$ is dimensionful for $p\neq 3$. Note also that from the perspective of supergravity $N$ needs to be an integer due to RR flux quantization through the $S^{8-p}$.

The background in \eqref{eq:10dsolR10} is asymptotically $\mathbb{R}^{1,9}$. To obtain a supergravity background suitable for holography one needs to take a near-horizon limit. As explained in~\cite{Itzhaki:1998dd}, this amounts to defining the dimensionless radial coordinate $U=r/(2\pi \ell_s)$ and taking the limit
\begin{equation}\label{eq:NHlimit}
\frac{g_sN U^{p-7}}{2\pi V_{6-p}}\gg 1\,.
\end{equation}
In this limit the supergravity background reduces to 
\begin{equation}\label{eq:NHsugra}
\begin{split}
ds^2_{10} &= (gU)^{\frac{7-p}{2}} ds^2_{p+1} + (gU)^{\frac{p-7}{2}}(dU^2 + U^2 d\Omega^2_{8-p})\,,\\
e^{\Phi}&= g_s (gU)^{\frac{(p-3)(7-p)}{4}}\,, \quad\qquad C_{p+1} = g_s^{-1}(gU)^{7-p} {\rm vol}_{p+1}\,,
\end{split}
\end{equation}
where we have defined $g$ through the relation
\begin{equation}\label{eq:gsugradef}
(2\pi \ell_s g)^{p-7} = \frac{g_s N}{2\pi V_{6-p}}\,. 
\end{equation}
The parameter $g$ is related to the coupling constant in a $(p+2)$-dimensional maximally supersymmetric gauged supergravity theory that can be used for an effective description of the dynamics of the D$p$-branes, see \cite{Boonstra:1998mp}. In the familiar example of D3-branes $1/g$ is proportional to the length scale $L$ of the AdS$_5 \times S^5$ solution. We note that the background~\eqref{eq:NHsugra} is not asymptotically locally AdS for $p\neq 3$.

To discuss the regimes of validity of the perturbative field theory and supergravity limits it is useful to follow~\cite{Itzhaki:1998dd} and define the effective coupling $g_{\rm eff}^2 = g^2_{\rm YM}N U^{p-3}$. The SYM theory is weakly coupled for $g_{\rm eff}^2 \ll 1$ which corresponds to the region $U \gg (g^2_{\rm YM} N)^{\frac{1}{3-p}}$ of the supergravity solution for $p<3$ and $U \ll (g^2_{\rm YM} N)^{\frac{1}{3-p}}$ for $p>3$. If one thinks of the radial coordinate $U$ of the supergravity solution in~\eqref{eq:NHsugra} as determining the energy scale in the dual field theory, then one finds agreement with the expectations from the dynamics of the SYM theory.  Indeed, for $p<3$ the SYM theory is asymptotically free and thus should be weakly coupled in the UV region described by large values of $U$. For $p=3$ the SYM theory is conformal and the YM coupling parametrizes a (subset) of the conformal manifold of $\mathcal{N}=4$ SYM. For $p>3$ the gauge theory is weakly coupled in the IR, i.e. for small values of $U$, but is not renormalizable and needs to be completed by additional physics in the UV. For $p=4$ the UV completion is in terms of the 6d $\mathcal{N}=(2,0)$ SCFT living on the worldvolume of coincident M5-branes. For $p=5$ we find the $\mathcal{N}=(1,1)$ little string theory in the UV. For $p=6$ there is no known UV completion without gravity and it is expected that one recovers the type IIA string theory in the UV description of D6-branes.\footnote{See \cite{Peet:1998wn,Bobev:2019bvq,Minahan:2022rsx} for a discussion of the somewhat exotic behavior of the 7d SYM theory that includes indications that the correct strongly coupled physics is not gravitational and may need to be described in terms of a negative coupling constant.\label{ft:D6branes}} The classical supergravity approximation is valid when both the curvature of the metric in~\eqref{eq:NHsugra} and the dilaton are small. This is obeyed in the regime $1\ll g_{\rm eff}^2 \ll N^{\frac{4}{7-p}}$, which implies that the supergravity background~\eqref{eq:NHsugra} can be trusted only for a range of values of the radial coordinate $U$. As is typical in holographic dualities, the classical supergravity regime does not overlap with the regime where perturbative field theory calculations are reliable. From here on, our discussion will be focused on the classical supergravity regime and the corresponding strongly coupled SYM dynamics.

%%%%%%
\subsection{Scaling similarity}
\label{subsec:ScalingSim}
%%%%%%
As described in \cite{Biggs:2023sqw}, the background in \eqref{eq:NHsugra} enjoys a scaling similarity. More concretely, if one rescales the coordinates in the $ds^2_{p+1} $ part of the metric as $x_{\mu} \to \gamma x_{\mu}$ and the radial coordinate as  $U \to \gamma^{-\frac{2}{5-p}} U$, then the classical supergravity action is rescaled as $S \to \gamma^{-\eta} S$, where the scaling exponent $\eta$ is given by\footnote{Many of the expressions below do not have a smooth $p=5$ limit. As explained in \cite{Bobev:2019bvq} the limit $p \to 5$ is subtle and should be treated with care by an appropriate renormalization of the gauge coupling. We will not dwell further on this subtlety here and as a result some of our results will appear to have singular behavior at $p=5$.}
\begin{equation}\label{eq:etadef}
\eta = \frac{(3-p)^2}{5-p}\,.
\end{equation}
This scaling transformation is not a symmetry of the supergravity background, nevertheless it has important consequences for physical observables. For instance, it determines that the on-shell action of the supergravity solution, which corresponds to the free energy in the dual SYM theory, should have the following scaling
\begin{equation}\label{eq:Sonshelscaling}
S_{\rm on-shell} \sim (g^{2}_{\rm YM} N)^{\frac{p-3}{5-p}} N^2\,.
\end{equation}
This scaling with the 't Hooft coupling of the free energy in the $(p+1)$-dimensional SYM theory is a non-trivial prediction for the strongly coupled dynamics of the planar gauge theory. One way to confirm this prediction with field theory methods is to employ supersymmetric localization. Indeed, one can place the maximally supersymmetric YM theory on $S^{p+1}$ and reduce the path integral to a matrix model \cite{Blau:2000xg,Minahan:2015jta}. The resulting matrix model can be analyzed in the large-$N$ and strong-coupling limit to find the following result for the free energy of the theory on $S^{p+1}$ of radius $\mathcal{R}$, see \cite{Bobev:2019bvq,Bobev:2024gqg},
\begin{equation}\label{eq:FSpsusyloc}
F_{S^{p+1}} = \mathfrak{c}_p \lambda^{\frac{p-3}{5-p}} N^2\,,
\end{equation}
where we have defined the dimensionless 't Hooft coupling $\lambda = g^2_{\rm YM} N \mathcal{R}^{3-p}$. The constant $\mathfrak{c}_p$ can be explicitly computed from the matrix model and takes the form
\begin{equation}\label{eq:cpdef}
\mathfrak{c}_p = \frac{2^{\frac{4p-16}{p-5}}\pi^{9+\frac{p}{2}+\frac{14}{p-5}}\left(\Gamma\left(\frac{5-p}{2}\right)\Gamma\left(\frac{7-p}{2}\right)\Gamma\left(\frac{p}{2}\right)\right)^{\frac{2}{5-p}}}{(p-7)(p-3)\Gamma\left(\frac{p-2}{2}\right)}\,.
\end{equation}
This result can be viewed as a quantum field theory derivation of the non-trivial power of $\lambda$ dictated by the scaling similarity of the dual supergravity solution. Moreover, the numerical coefficient $\mathfrak{c}_p$ agrees with the on-shell action of the spherical D$p$-brane solutions of \cite{Bobev:2018ugk}.\footnote{The spherical brane solutions share the same large $U$ asymptotics with the background in \eqref{eq:NHsugra} but are different otherwise. The characteristic scale $\mathcal{R}$ set by the radius of the spherical D$p$-brane acts as an IR cutoff of the dynamics which is manifested in the bulk as a smooth cap-off of the geometry for small values of $U$, see~\cite{Bobev:2018ugk} for more details.} This is a rare example of a precise agreement between observables computed on the field theory and gravity sides of this non-conformal holographic duality.

%%%%%%%%%%%%%%
\subsection{Auxiliary AdS}
\label{subsec:AuxAdS}
%%%%%%%%%%%%%%

As explained in \cite{Kanitscheider:2008kd,Kanitscheider:2009as,Biggs:2023sqw}, the string frame supergravity backgrounds \eqref{eq:NHsugra} can be thought of as a formal dimensional reduction of a simple AdS$_{p+2+\eta}\times S^{8-p}$ background with constant $(8-p)$-form flux on the $S^{8-p}$, see Appendix A of \cite{Biggs:2023sqw} for more details. Importantly, the $(10+\eta)$-dimensional metric has the standard product form and the length scales of the AdS$_{p+2+\eta}$ and $S^{8-p}$ factors are related as $\frac{L_{S^{8-p}}}{L_{{\rm AdS}_{p+2+\eta}}} = \frac{7-p}{p+1+\eta}$. This formal dimensional uplift geometrizes the running dilaton of the 10d supergravity backgrounds \eqref{eq:NHsugra} and, as we discuss below, is a very useful device for holographic calculations. The fact that the uplifted background has an AdS factor in the metric is closely related to the scaling similarity of the near-horizon limit of the D$p$-brane supergravity backgrounds~\eqref{eq:NHsugra}. We note that, when implementing the uplift to $10+\eta$ dimensions, it is important to first use Hodge duality to convert the ``electric'' RR potential $C_{p+1}$ to ``magnetic'' flux on the $S^{8-p}$ part of the 10d supergravity background.

In general, we will consider the uplift to a $10+\eta$ background with an AdS$_{p+2+\eta}$ factor as a mathematical trick that helps us streamline and organize holographic calculations. For some values of $p$ however, there is a suggestive physical picture that may explain the existence of this structure. For $p=1$ and $p=4$ one finds that $\eta=1$ and therefore the uplifted solution has an integer-dimensional AdS factor. This can perhaps be explained as follows. For $p=1$ one can S-dualize the D1-branes sourcing the supergravity background to F1 fundamental strings which, after using T-duality, can be thought of as a dimensional reduction of M2-branes. These M2-branes, in an appropriate limit, can then produce an AdS$_4$ vacuum of 11d supergravity. For $p=4$ we have D4-branes in type IIA string theory which at strong coupling uplift to M5-branes in 11d. The low-energy dynamics of coincident M5-branes is described by a 6d $\mathcal{N}=(2,0)$ SCFT dual to an AdS$_7$ vacuum of 11d supergravity. We stress that the reasoning and chain of dualities described above do not constitute a rigorous derivation and provide only a suggestive explanation of the auxiliary AdS$_{4,7}$ structure for $p=1,4$. For $p=\{-1,0,2\}$ we find fractional values of $\eta$ and these cases can be considered as generic in the holographic discussion below.\footnote{As discussed above, the case $p=5$ is more subtle and cannot be obtained as a smooth limit of this general discussion. For $p=6$ we have $\eta = -9$ and therefore we find a formal dimensional \textit{reduction} from 10d to 1d and an AdS space of negative dimension. It is unclear how to think of this exotic structure which may be related to the unusual features of the strong-coupling limit of the 7d SYM theory mentioned in Footnote~\ref{ft:D6branes}. }

The utility of the auxiliary AdS structure described above is very clear in the calculation of holographic 2pt-functions of supergravity fields. One can take any perturbative excitation\footnote{We focus on perturbations that are scalars on the $ds^2_{p+1}$ part of the space-time, i.e. supergravity modes dual to single-trace scalar operators in the SYM gauge theory.} of the supergravity fields around the background in~\eqref{eq:NHsugra} and then formally uplift it to a perturbation of the AdS$_{2+p+\eta}\times S^{8-p}$ background. Employing the standard holographic calculation of 2pt-functions, we then conclude that this mode should behave like a scalar mode on AdS$_{p+2+\eta}$ dual to an operator in an auxiliary CFT in $p+1+\eta$ dimensions. The mass of the mode $m$ is related to the auxiliary conformal dimension $\Delta$ by the standard formula
\begin{equation}\label{eq:massDeltarel}
m^2L_{{\rm AdS}_{p+2+\eta}}^2 = \Delta(\Delta-(p+1+\eta))\,.
\end{equation}
Reducing back to the $p+1$-dimensional physical space-time on which the SYM gauge theory lives, one finds that the scalar operators in the strongly coupled planar limit of the gauge theory should have the following 2pt-function
\begin{equation}\label{eq:2ptDpschem}
\langle \O(x)\O(y) \rangle \sim \frac{1}{|x-y|^{2\Delta-\eta}}\,.
\end{equation}
This observation, discussed in \cite{Biggs:2023sqw} (see also \cite{Jevicki:1998ub,Sekino:1999av,Kanitscheider:2008kd,Kanitscheider:2009as}), provides an efficient way to calculate the holographic 2pt-functions for the non-conformal D$p$-brane setup and strongly suggests a method to compute higher-point correlation functions. Presenting the detailed implementation of this idea will be the main subject of Section~\ref{sec:intCFT}.

The result in \eqref{eq:2ptDpschem}, together with \eqref{eq:massDeltarel}, reduces the calculation of the holographic D$p$-brane 2pt-functions of supergravity modes to the calculation of the spectrum of masses for the infinite tower of KK modes around the supergravity backgrounds in~\eqref{eq:NHsugra}. This complicated technical problem was solved in \cite{Kim:1985ez} for the well-known AdS$_5\times S^5$ solution but, to the best of our knowledge, has not been solved in full generality for the $p\neq 3$ backgrounds in~\eqref{eq:NHsugra}. Nevertheless, some partial results are available in the literature. As discussed in \cite{Biggs:2023sqw}, see also \cite{Sekino:1999av}, we have the following sets of operator dimensions arising from some of the bosonic KK modes on $S^{8-p}$
\begin{equation}\label{eq:DeltaKK}
\Delta = \frac{7-p}{5-p} (b+2) + \frac{2}{5-p}\ell\,.
\end{equation}
The parameter $\ell$ is an integer that labels the KK level, while the parameter $b$ depends on the particular supergravity mode. For instance, for KK modes dual to BPS operators of the schematic form ${\rm Tr}[\phi^{\ell}]$ in the $\ell$-symmetric traceless representation of the ${\rm SO}(9-p)$ R-symmetry of the gauge theory, we have $b=-2$ and $\ell \geq 2$, while for operators with spin of the schematic form ${\rm Tr}[\partial_{\mu}\phi\phi^{\ell}]$ we have $b=-1$ and $\ell \geq 2$. Here $\phi$ schematically denotes the $(9-p)$ scalar fields in the SYM theory that transform in the fundamental representation of the R-symmetry and the adjoint representation of the gauge group.

As we discuss in Section~\ref{sec:intCFT}, if one is to exploit the power of the auxiliary AdS structure to compute holographic 3pt-correlation functions for D$p$-branes, one needs the explicit form of the cubic supergravity couplings of the KK modes. Deriving this cubic effective action by a direct calculation in 10d supergravity is technically involved even for the AdS$_5 \times S^5$ background, see~\cite{Arutyunov:1999en}, and has not been performed for the D$p$-brane backgrounds in~\eqref{eq:NHsugra}. This is an important open problem that we leave for the future.

The discussion so far was based on 10d IIA/B string theory and its low-energy supergravity limit. It is well-known however, that for any $p$ there is a consistent truncation of the near-horizon D$p$-brane backgrounds in \eqref{eq:NHsugra} to solutions of an appropriate maximally supersymmetric gauged supergravity theory in $p+2$ dimensions \cite{Boonstra:1998mp}. The relevant gauged supergravity theories for $0 \leq p \leq 6$ were summarized in \cite{Bobev:2018ugk,Bobev:2024gqg}, while the theory for $p=-1$ was recently discussed in \cite{Ciceri:2025maa}. This lower-dimensional gauged supergravity theory provides access only to a finite subset of the KK supergravity modes but describes a fully non-linear action for them. This truncated $(p+2)$-dimensional action can be used in particular to efficiently calculate, amongst other things, the masses and cubic bulk couplings needed for the holographic evaluation of 2pt- and 3pt-functions; we will discuss this calculation elsewhere~\cite{BBGMP}. The auxiliary AdS structure discussed above suggests that these $(p+2)$-dimensional gauged supergravity theories, which do not have AdS vacua, can be thought of as non-linear theories in $p+2+\eta$ dimensions that do have AdS solutions.

%%%%%%%%%%%
\section{Correlation functions for non-conformal branes}
\label{sec:intCFT}
%%%%%%%%%%%

The auxiliary AdS$_{p+2+\eta}$ structure described above suggests a holographic prescription for the computation of correlation functions of operators in the $(p+1)$-dimensional maximally supersymmetric YM theory. One starts with a CFT correlation function in an auxiliary $D$-dimensional Euclidean space $\mathbb{R}^{D}$ with $D=p+1+\eta$, whose position dependence is constrained by conformal invariance. One then integrates this correlation function over the positions $\xi_{x_i}$ in the $\mathbb{R}^{\eta}$ subspace of $\R^D$. Denoting the positions of the operators on $\mathbb{R}^{D}$ by $X_i$ and those on $\mathbb{R}^{p+1}$ by $x_i$, this prescription can be summarized concisely by the following expression\footnote{For $p=-1$ the brane world-volume is a point and therefore there is no position dependence of the correlation functions. The results for $n$pt-functions obtained by our method then reduce to some numbers. It is unclear how these numerical values should be interpreted, especially given that we do not have a reliable method to determine the normalization of holographic correlators in our setup. For this reason we will assume that $p\geq 0$ in the discussion below.}
\begin{empheq}[box=\fbox]{equation}
\begin{aligned}\label{eq:nptprescr}
	~\langle\O_1(x_1)\ldots\O_n(x_n)\rangle_{\eta} = \int\prod_{i=1}^n d^{\eta}\xi_{x_i}\langle\O_1(X_1)\ldots\O_n(X_n)\rangle\,.~
\end{aligned}
\end{empheq}
The subscript $\eta$ on the LHS indicates the fractional-dimensional holographic origin of the correlation function. While we do not have a derivation of this prescription for the calculation of holographic correlators, the structure of the D$p$-brane supergravity backgrounds~\eqref{eq:NHsugra} suggests that it is valid at the two-derivative classical supergravity approximation and for single-trace local operators in the SYM gauge theory dual to supergravity KK modes. It is of course very important to derive this prescription by a careful bulk analysis of holographic correlation functions using the analog of Witten diagrams, as initiated in \cite{Kanitscheider:2008kd,Kanitscheider:2009as}.

The structure of the CFT $n$pt-functions appearing on the RHS of \eqref{eq:nptprescr} is not completely determined by the conformal symmetry for $n\geq4$ and depends on the dynamics of the auxiliary $D$-dimensional CFT. Since we have no knowledge about this putative CFT dynamics we will restrict ourselves to correlators with $n=1,2,3$ whose position dependence is completely determined by conformal invariance. In the simplest case of $n=1$, we already find the non-trivial result that the corresponding 1pt-functions in the SYM gauge theory should vanish. This result is compatible with the explicit calculations of some holographic 1pt-functions in~\cite{Kanitscheider:2008kd,Kanitscheider:2009as} which were computed at finite temperature and vanish in the supersymmetric limit $T \to 0$. It will be interesting to establish the vanishing of 1pt-functions of all KK modes in the background~\eqref{eq:NHsugra} more rigorously. We hasten to add that this vanishing of the 1pt-functions, as well as the results for the 2pt- and 3pt-functions discussed below, are only valid for the ${\rm SO}(9-p)$ invariant vacuum of the SYM theory and do not apply in general for other vacua. This includes the ones discussed in \cite{Berenstein:2002jq} for $p=0$, which preserve 16 supercharges.

Before we proceed with the implementation of the prescription in \eqref{eq:nptprescr} for $n=2$ and $n=3$, it is useful to understand how 2pt- and 3pt-functions of single-trace operators dual to supergravity KK modes scale with $N$ and the dimensionful 't Hooft coupling $g_{\rm YM}^2N$. To this end we can appeal to the standard AdS/CFT logic. In the well-studied $p=3$ example of AdS/CFT the type IIB supergravity action has a prefactor $\zeta$ that scales as $\zeta \sim N^2$. The 2pt-functions of single-trace BPS operators in the 4d $\mathcal{N}=4$ SYM dual to the supergravity KK modes then scale as $\langle \O \O\rangle \sim \zeta^0$, while the 3pt-functions scale as $\langle \O \O \O\rangle\sim \zeta^{-\frac{1}{2}}$.\footnote{The scaling will be different for multi-trace operators or for heavy local operators corresponding to string or D-brane excitations.} For the non-conformal setup with $p\neq 3$ we need to change $\zeta$ to $\zeta_{p} \sim (g_{\rm YM}^2N)^{\frac{(p-3)}{(5-p)}} N^2$, as dictated by the supergravity action scaling in \eqref{eq:Sonshelscaling}. This in turn will determine the scaling of 2pt- and 3pt-functions as follows
\begin{equation}\label{eq:zetascaling}
\langle \O \O\rangle \sim \zeta_p^0\,, \qquad\qquad \langle \O \O \O\rangle\sim \zeta_p^{-\frac{1}{2}}\,.
\end{equation}
It would be nice to confirm this scaling with an explicit evaluation of the bulk Witten diagrams. Note that since $g_{\rm YM}$ is dimensionful for $p\neq 3$ the 3pt-functions will change with the energy scale in a way dictated by the RG running of the coupling.

%%%%%%%%%%%%%%
\subsection{2pt-functions}
\label{subsec:2pt}
%%%%%%%%%%%%%%
As motivated above, we start with the standard expression for the 2pt-function of a CFT in $D$ dimensions and integrate it over the auxiliary $\eta$-dimensional subspace. We denote the positions in the auxiliary space by capital letters, i.e. $X,Y\in\R^{D}$. We define $d=p+1$ for notational simplicity and split the coordinates $X=(x,\xi_x)$, with $x\in\R^{d}$ and $\xi_x\in\R^\eta$, and similarly for $Y$. The desired $d$-dimensional 2pt-function $\langle\O(x)\O(y)\rangle_{\eta}$ is then computed by the integral
\begin{align}\label{eq:2pt_int}
	\langle\O(x)\O(y)\rangle_{\eta} = \int d^{\eta}\xi_x d^{\eta}\xi_y~ \frac{1}{|X-Y|^{2\Delta}} \,,
\end{align}
where $\Delta$ is the conformal dimension of the operator $\O$ in the auxiliary CFT$_D$. To evaluate this integral, we split the coordinates as described above to obtain
\begin{equation}
\begin{split}
    \langle\O(x)\O(y)\rangle_{\eta} = \int d^{\eta}\xi_x d^{\eta}\xi_y~ \frac{1}{\big(|x-y|^2 + |\xi_x - \xi_y|^2\big)^\Delta} \,.
\end{split}
\end{equation}
Shifting the integration variable $\xi_y\mapsto\xi_y+\xi_x$ decouples the two integrals, and after performing the $\xi_y$ integration we find
\begin{align}\label{eq:2ptreseta}
	\langle\O(x)\O(y)\rangle_{\eta} = \left(\int d^\eta\xi_x\right)\times \frac{\pi^{\eta/2}\,\Gamma(\Delta-\etah)}{\Gamma(\Delta)} \frac{1}{|x-y|^{2\Delta-\eta}}\,.
\end{align}
Note that convergence of the $\xi_y$ integral over $\R^\eta$ requires $\Delta>\etah$. This condition is indeed obeyed by all modes in the supergravity KK spectrum~\eqref{eq:DeltaKK}, except a single one at $p=0$ with $b=-2$ and $\ell=2$. As noted in \cite{Biggs:2023sqw} and discussed around \eqref{eq:2ptDpschem} above, the spacetime dependence of the result \eqref{eq:2ptreseta} takes the usual form of a CFT 2pt-function but with a shifted conformal dimension $\Delta_\eta=\Delta-\etah$. The remaining $\xi_x$ integral evaluates to the volume of $\R^\eta$, yielding an infinite factor which we denote by
\begin{equation}\label{eq:Vetadef}
    \Veta \equiv \text{Vol}(\R^\eta) = \int d^\eta \xi\,.
\end{equation}
Given the ad hoc nature of our procedure we do not have a watertight method to regulate this divergent prefactor. It may reflect an ambiguity in the normalization of the 2pt-function intrinsic to our procedure, or may have a more physical origin. For instance, the running of the YM coupling in the $(p+1)$-dimensional gauge theory should set an RG energy scale for all correlation functions, which in turn should be related to the coordinate $U$ in the dual supergravity solution~\eqref{eq:NHsugra}. There is a potential ambiguity in this holographic map of energy scales that may be used to determine the correct procedure to fix the value of $\Veta$.

While we will not commit to a specific value of $\Veta$, we note that there is a simple modification of the procedure above that leads to a finite value of the 2pt-function. We can first use the $D$-dimensional translational invariance of the CFT to set $Y=0$ in \eqref{eq:2pt_int} and thus eschew the need to integrate over $\xi_y$. Performing the integral over $\xi_x$, we then find the following finite answer
\begin{equation}\label{eq:2ptY0}
	\langle\O(x)\O(0)\rangle_{\eta} = \frac{\pi^{\eta/2}\, \Gamma(\Delta-\eta/2)}{\Gamma(\Delta)} \frac{1}{|x|^{2\Delta-\eta}}\,.
\end{equation}
We can now appeal to the Poincar\'e invariance of the SYM vacuum and shift $x\mapsto x-y$ in~\eqref{eq:2ptY0} to restore the dependence of the 2pt-functions on $y$. This sleight of hand amounts to setting $\Veta=1$ in \eqref{eq:2ptreseta}.

Some comments on the normalization of the 2pt-functions are in order. We start with a unit normalized CFT$_D$ 2pt-functions in \eqref{eq:2pt_int}. This then leads to the $d$-dimensional SYM 2pt-function normalization in~\eqref{eq:2ptreseta} and~\eqref{eq:2ptY0}.  Presumably the careful bulk supergravity calculation of this 2pt-function in the background~\eqref{eq:NHsugra} will fix a preferred normalization which will be nice to determine and compare with a dual field theory calculation. We note in passing that the 2pt-function normalization in \eqref{eq:2ptY0} is tantalizingly close to the inverse of the normalization of the 2pt-function in standard AdS/CFT bulk calculations, see for instance Equation (17) in~\cite{Freedman:1998tz}. This may be pure coincidence or a hint on how to determine the correct bulk supergravity normalization.

Finally, we note that the 2pt-functions of some of the operators with dimensions~\eqref{eq:DeltaKK} for $p=0$ have been computed using lattice methods, see \cite{Hanada:2011fq}. Interestingly, the power law behavior of the 2pt-function in \eqref{eq:2ptY0} agrees with the lattice simulations for relatively low values of the rank of the gauge group $N$.

%%%%%%%%%
\subsection{3pt-functions}
\label{subsec:3pt}
%%%%%%%%%
We now apply the procedure in~\eqref{eq:nptprescr} to the 3pt-function. In the $D$-dimensional CFT we have the 3pt-function
\begin{equation}\label{eq:3ptDdim}
\langle\O_1(X)\O_2(Y)\O_3(Z)\rangle = \frac{1}{|X-Y|^{2\Theta_3}|X-Z|^{2\Theta_2}|Y-Z|^{2\Theta_1}}\,,
\end{equation}
where the scalar operators $\O_i$ have conformal dimension $\Delta_i$ and we introduced the notation
\begin{equation}\label{eq:Thetai_def}
    \Theta_i \equiv \frac{1}{2}(\Delta_1 + \Delta_2 + \Delta_3) - \Delta_i\,.
\end{equation}
Following \eqref{eq:nptprescr}, the $d$-dimensional SYM holographic 3pt-function is computed by the integral
\begin{align}\label{eq:3pt_int}
	\langle\O_1(x)\O_2(y)\O_3(z)\rangle_\eta = \int d^{\eta}\xi_x d^{\eta}\xi_y d^{\eta}\xi_z~\frac{1}{|X-Y|^{2\Theta_3}|X-Z|^{2\Theta_2}|Y-Z|^{2\Theta_1}}\,.
\end{align}
To simplify the notation in the calculation of this integral we define the $d$-dimensional distances
\begin{equation}
{\rm d}_{xy} = |x-y|\,, \quad {\rm d}_{xz} = |x-z|\,, \quad {\rm d}_{yz} = |y-z|\,.  
\end{equation}
We now proceed to explicitly evaluate the integral in \eqref{eq:3pt_int}. Note that one can again decouple one of the integrals by shifting the integration variables, e.g. $\xi_x\mapsto\xi_x+\xi_z$, $\xi_y\mapsto\xi_y+\xi_z$, to obtain
\begin{align}\label{eq:3ptetaintdef}
	\langle\O_1(x)\O_2(y)\O_3(z)\rangle_\eta = \Veta \cdot \int \frac{d^{\eta}\xi_x d^{\eta}\xi_y}{({\rm d}_{xy}^2+|\xi_x-\xi_y|^2)^{\Theta_3}({\rm d}_{xz}^2+|\xi_x|^2)^{\Theta_2}({\rm d}_{yz}^2+|\xi_y|^2)^{\Theta_1}}\,.
\end{align}
To compute the remaining double integral, we will draw inspiration from the results in \cite{Davydychev:1992mt} where similar integrals were encountered in the calculation of two-loop Feynman diagrams in perturbative QFT. We proceed by employing the Mellin-Barnes representation by making use of the integral identity
\begin{equation}
	\frac{1}{(S+T)^{\nu}} = \frac{1}{\Gamma(\nu)}~\frac{1}{2\pi {\rm i}} \int_{-\ri\infty}^{+\ri\infty} du \, \Gamma(\nu+u)\Gamma(-u)\,\frac{T^u}{S^{\nu+u}}\,.
\end{equation}
Applying this identity twice to the terms with powers $\Theta_{2}$ and $\Theta_3$ in the denominator of \eqref{eq:3ptetaintdef} we arrive at
\begin{align}
\begin{split}
	\langle\O_1(x)\O_2(y)\O_3(z)\rangle_\eta &= \frac{\Veta}{\Gamma(\Theta_2)\Gamma(\Theta_3)}~\frac{1}{(2\pi\ri)^2} \int_{-\ri\infty}^{+\ri\infty}dudv\,(\rd_{xy}^2)^u (\rd_{xz}^2)^v\Gamma(-u) \Gamma(-v)\\
	&\qquad\qquad\qquad\qquad\qquad\qquad\qquad\times\Gamma(\Theta_3+u)\Gamma(\Theta_2+v)\,\I(u,v)\,,
\end{split}
\end{align}
with $\I(u,v)$ given by
\begin{align}
	\I(u,v) = \int\frac{d^\eta\xi_xd^\eta\xi_y}{(\xi_x^2)^{\Theta_2+v}~[(\xi_x-\xi_y)^2]^{\Theta_3+u}~(\rd_{yz}^2+\xi_y^2)^{\Theta_1}}\,.
\end{align}
Integrals of this type are related to two-loop diagrams in perturbative QFT and can be evaluated by using the standard Feynman parametrization.\footnote{Alternatively, a version of this integral relevant for Lorentzian space QFTs can be found in textbooks, see for instance Equation (10.39) in \cite{Smirnov:2012gma}.} Explicitly, we find
\begin{align}
	\I(u,v)=\frac{\pi^{\eta}}{(\rd_{yz}^2)^{\Theta+u+v-\eta}}\,\frac{\Gamma(\etah-\Theta_2-v)\Gamma(\etah-\Theta_3-u)\Gamma(\Theta_2+\Theta_3+u+v-\etah)\Gamma(\Theta+u+v-\eta)}{\Gamma(\Theta_2+v)\Gamma(\Theta_3+u)\Gamma(\Theta_1)\Gamma(\etah)}\,,
\end{align}
where we used the definition
\begin{equation}
\Theta \equiv \Theta_1+\Theta_2+\Theta_3 = \frac{1}{2}\left(\Delta_1+\Delta_2+\Delta_3\right)\,.
\end{equation}
We have therefore reduced the 3pt-function integral \eqref{eq:3pt_int} to the following two-fold Mellin-Barnes representation:\footnote{Note that, as usual, the integration contour is chosen to separate the left- and right-moving poles, which for generic values of $\eta$ and the $\Theta_i$ can be easily achieved. Moreover, there is direct evidence from the spectrum of operator dimensions corresponding to supergravity KK modes \eqref{eq:DeltaKK} that for general KK modes no problematic overlap of poles occurs. We will further comment on this below~\eqref{eq:3pt_Appell}.}
\begin{align}\label{eq:Mellin_int}
\begin{split}
	\langle\O_1(x)\O_2(y)\O_3(z)\rangle_\eta &= \frac{\N_{\Delta_i}}{(2\pi\ri)^2}\int_{-\ri\infty}^{+\ri\infty}dudv\,\frac{(\rd_{xy}^2)^u(\rd_{xz}^2)^v}{(\rd_{yz}^2)^{u+v+\Theta-\eta}}\\
	&\qquad\qquad\qquad\qquad~\times\Gamma(-u)\Gamma(\tfrac{\eta}{2}-\Theta_3-u)\Gamma(-v)\Gamma(\tfrac{\eta}{2}-\Theta_2-v)\\
	&\qquad\qquad\qquad\qquad~\times\Gamma(\Theta_2+\Theta_3+u+v-\tfrac{\eta}{2})\Gamma(\Theta+u+v-\eta)\,,
\end{split}
\end{align}
where the normalization factor $\N_{\Delta_i}$ is given by
\begin{align}\label{eq:normfactor}
	\N_{\Delta_i}=\frac{\pi^\eta\,\Veta}{\Gamma(\Theta_1)\Gamma(\Theta_2)\Gamma(\Theta_3)\Gamma(\etah)}\,.
\end{align}
We can evaluate the above Mellin-Barnes integral by closing the contours of integration to the right. There are two infinite sequences of poles in both $u$ and $v$, coming from the gamma functions in the second line of \eqref{eq:Mellin_int}. The poles in $u$ are located at $u=n_1$ and $u=\etah-\Theta_3+n_2$, and similarly the poles in $v$ are at $v=m_1$ and $v=\etah-\Theta_2+m_2$ with $\{n_{1,2},m_{1,2}\}\in \mathbb{N}_{0}$. In the following, we assume that the two series of poles for $u$ and $v$ do not overlap, which amounts to the condition that
\begin{align}\label{eq:non_overlapping_assumption}
	\tfrac{\eta}{2}-\Theta_2\,,\,\tfrac{\eta}{2}-\Theta_3 ~\notin\, \mathbb{Z}\,.
\end{align}
This assumption is physically motivated by recalling the spectrum of operator dimensions~\eqref{eq:DeltaKK}. Some experimentation shows that the typical situation is that the above condition is indeed satisfied, c.f. the more detailed discussion in the last paragraph below equation \eqref{eq:3pt_Appell}.\footnote{For the special cases when it is not, one needs to revisit the analysis which follows. In particular, overlapping poles in \eqref{eq:Mellin_int} will lead to explicit logarithmic terms in $A$ and/or $B$.}

Then, provided that \eqref{eq:non_overlapping_assumption} holds, taking the residues of the two series of poles in $u$ and $v$ gives rise to four different infinite series in the variables $A$ and $B$ defined as
\begin{align}\label{eq:def_AB}
	A=\frac{\rd_{xy}^2}{\rd_{yz}^2}\,,\qquad B=\frac{\rd_{xz}^2}{\rd_{yz}^2}\,.
\end{align}
As noticed in \cite{Davydychev:1992mt}, these series can be written in terms of Appell hypergeometric functions of two variables,
\begin{align}\label{eq:Appell}
	F_4(a,b;c,d;A,B) = \sum_{m=0}^\infty\sum_{n=0}^\infty \frac{(a)_{m+n}(b)_{m+n}}{(c)_m(d)_n}\cdot\frac{A^m B^n}{m!\,n!}\,,
\end{align}
where $(a)_k$ denotes the usual Pochhammer symbol. The Appell hypergeometric function is a two-variable generalization of the usual hypergeometric function. The region of convergence of the series expansion~\eqref{eq:Appell} is $\sqrt{|A|}+\sqrt{|B|}<1$, and outside this region the function can be defined by analytic continuation, see for example~\cite{Exton}. With the definition of $A$ and $B$ given in~\eqref{eq:def_AB}, the triangle inequality $\rd_{xy}+\rd_{xz}\geq\rd_{yz}$ implies that $\sqrt{|A|}+\sqrt{|B|}\geq 1$, meaning that any three points $\{x,y,z\}$ on $\R^d$ will give $(A,B)$ outside of the region of convergence of the series expansion~\eqref{eq:Appell}; and thus an analytic continuation of the Appell $F_4$ function is required. In Appendix~\ref{sec:Appendix} we discuss this analytic continuation in some detail together with its numerical implementation, which allows us to evaluate the $F_4$ function in the physical region of interest. We also note that for $p=0$ the points $\{x,y,z\}$ lie on a line which then leads to $\sqrt{|A|}+\sqrt{|B|}=1$. This case, as well as the collinear arrangement of three points for $p>0$, is discussed further in Appendix~\ref{sec:Appendix}.

Using the results above we arrive at the following final expression for the 3pt-function:
\begin{align}\label{eq:3pt_Appell}
\begin{split}
	\langle\O_1(x)\O_2(y)\O_3(z)\rangle_\eta &= \N_{\Delta_i}(\rd_{yz}^2)^{\eta-\Theta}\times\\[3pt]
%%%
	&\qquad\quad\Big[\Gamma(\tfrac{\eta}{2}-\Theta_2)\Gamma(\tfrac{\eta}{2}-\Theta_3)\Gamma(\Theta-\eta)\Gamma(\Theta_2+\Theta_3-\tfrac{\eta}{2})\,\times\\
	&\qquad\qquad F_4(\Theta-\eta,\Theta_2+\Theta_3-\tfrac{\eta}{2};1+\Theta_3-\tfrac{\eta}{2},1+\Theta_2-\tfrac{\eta}{2}; A,B)\\[3pt]
%%%
	&\qquad~~+B^{\etah-\Theta_2}\Gamma(\Theta_2-\etah)\Gamma(\etah-\Theta_3)\Gamma(\Theta_3)\Gamma(\Theta_1+\Theta_3-\etah)\,\times\\
	&\qquad\qquad F_4(\Theta_3,\Theta_1+\Theta_3-\etah;1+\Theta_3-\etah,1-\Theta_2+\etah; A,B)\\[3pt]
%%%
	&\qquad~~+A^{\etah-\Theta_3}\Gamma(\etah-\Theta_2)\Gamma(\Theta_3-\etah)\Gamma(\Theta_2)\Gamma(\Theta_1+\Theta_2-\etah)\,\times\\
	&\qquad\qquad F_4(\Theta_2,\Theta_1+\Theta_2-\etah;1-\Theta_3+\etah,1+\Theta_2-\etah; A,B)\\[3pt]
%%%
	&\qquad~~+A^{\etah-\Theta_3}B^{\etah-\Theta_2}\Gamma(\Theta_2-\etah)\Gamma(\Theta_3-\etah)\Gamma(\Theta_1)\Gamma(\etah)\,\times\\
	&\qquad\qquad F_4(\Theta_1,\etah;1-\Theta_3+\etah,1-\Theta_2+\etah; A,B)\Big].
\end{split}
\end{align}
This is our main result and it constitutes a non-trivial prediction about the kinematical structure of the holographic 3pt-functions for non-conformal D$p$-branes. Clearly, the position dependence of the 3pt-function is not compatible with conformal invariance, but the answer is manifestly Poincar\'e invariant since it depends only on the magnitude of the relative distances $\rd_{xy}^2$, $\rd_{xz}^2$, $\rd_{yz}^2$ through the ratios $A$ and $B$. Given the non-trivial structure of the answer it is clearly desirable to derive this result by other methods and scrutinize it. To this end, in Section~\ref{subsec:consistency} below we subject the expression~\eqref{eq:3pt_Appell} to several consistency checks.

Note that $A$ and $B$ are invariant under rescaling which means that the expression in the square bracket in~\eqref{eq:3pt_Appell} is scale invariant. This in turn implies that both the 2pt- and 3pt-functions in~\eqref{eq:2ptreseta} and~\eqref{eq:3pt_Appell} are scale covariant if one assigns a scaling weight~$\Delta_{\eta}=\Delta-\eta/2$ to every operator and takes into account that the prefactor $\zeta_{p}$ discussed around~\eqref{eq:zetascaling} is dimensionful and $(\zeta_p)^{\eta}$ has units of length for any value of $p$. This may naively suggest that we have a scale covariant theory which is not conformally invariant. It was noted however in~\cite{Hanada:2011fq,Biggs:2023sqw} that massive string states have exponentially decaying 2pt-functions and therefore the scale covariance of correlation functions must be a property only of the operators in the SYM theory that are dual to supergravity KK modes.

Similarly to the expression for the 2pt-function~\eqref{eq:2ptreseta}, the 3pt-function suffers from an ambiguity stemming from the factor of $\Veta$ in the normalization $\N_{\Delta_i}$ given in \eqref{eq:normfactor}. As discussed below \eqref{eq:Vetadef}, we will not commit to a specific regularization of this infinite volume factor. As in the case of the 2pt-function, see \eqref{eq:2ptY0}, to find a finite value of the correlator we can fix one of the points on $\mathbb{R}^D$, say $Z=0$ and then do not integrate over $\xi_z$. This will effectively result in setting $\Veta=1$ in~\eqref{eq:normfactor}. Importantly, the 3pt-function~\eqref{eq:3pt_Appell} should be properly normalized by including a finite coefficient determined by the cubic couplings in the effective supergravity action for the KK modes around the background~\eqref{eq:NHsugra}. These cubic couplings are not available in the literature and it will be most interesting to derive them and combine the result with~\eqref{eq:3pt_Appell} to find the complete holographic 3pt-function in the dual SYM gauge theory. 

The final result for the 3pt-function~\eqref{eq:3pt_Appell} appears ill-defined for some special values of the parameters. Indeed, we need to impose the condition in~\eqref{eq:non_overlapping_assumption}, which was assumed in the derivation of~\eqref{eq:3pt_Appell}, to ensure that some of the gamma functions in~\eqref{eq:3pt_Appell} do not diverge. There are additional potential divergences from the gamma functions of the type $\Gamma(\Theta_i+\Theta_j-\etah)$ or $\Gamma(\Theta-\eta)$.\footnote{Note that the explicit factors of $\Gamma(\Theta_i)$ appearing in \eqref{eq:3pt_Appell} are cancelled by the normalization factor $\N_{\Delta_i}$, c.f.~\eqref{eq:normfactor}, and hence these gamma functions do not lead to any divergences in the 3pt-function. This will become important when we consider the extremal limit $\Theta_i\to0$, see the discussion around equation \eqref{eq:3pt_extremal}.} However, we can use physical arguments based on unitarity, the operator spectrum of KK modes, and the allowed arrangements of operator dimensions in a 3pt-functions, to argue that these divergences are not realized.

First, note that the unitarity bound for scalar operators in the auxiliary CFT$_D$ implies $\Delta_i\geq\frac{d-2}{2}+\etah$. Using this we find that $\Theta>\eta$ for all $p>-1$. Alternatively, one can use the explicit values for the dimensions in~\eqref{eq:DeltaKK} to see that the same condition is obeyed. This ensures that there is no divergence from the term with $\Gamma(\Theta-\eta)$ in~\eqref{eq:3pt_Appell}. Second, note that the~${\rm SO}(9-p)$ selection rules, which organize the spectrum of KK-modes, will severely constrain the allowed supergravity cubic couplings and thus exclude some possible arrangements of $\Delta_i$ in 3pt-functions. In particular, as observed in all maximally supersymmetric examples of traditional AdS/CFT, only sub-extremal cubic couplings are non-vanishing, i.e. couplings between three operators with dimensions $(\Delta_i,\Delta_j,\Delta_k)$ obeying $\Delta_i+\Delta_j>\Delta_k$ for all permutations. Indeed, in the D$p$-brane setup one can also derive this condition by using ${\rm SO(9-p)}$ R-symmetry selection rules which imply that the 3pt-function of operators $\mathcal{O}_i$ in the $\ell_i$ symmetric traceless  representation can be non-zero only if $\ell_i+\ell_j \geq \ell_k$ and $\ell_i+\ell_j+\ell_k = 2 n$ where $n$ is a positive integer. In our setting, after using the dimensions in~\eqref{eq:DeltaKK} with $b=-2$, this implies that $\Theta_i>0$ for all $i$. We note in passing that this argument avoids any subtleties with terms involving a factor of $\Gamma(\Theta_i)$ in \eqref{eq:3pt_Appell} and \eqref{eq:normfactor}. However, the conditions above are not enough to exclude potential divergences when $\Theta_i+\Theta_j=\etah$. To argue against such a scenario, we appeal to the explicit spectrum of operator dimensions of KK modes in~\eqref{eq:DeltaKK}. We find that the only mode that violates the condition $\Theta_i+\Theta_j > \etah$ is the KK mode for $p=0$, $b=-2$, and $\ell=2$.\footnote{Recall that this is precisely the mode for which the 2pt-function in \eqref{eq:2ptreseta} exhibits a divergence. This is not an accident, since from \eqref{eq:Thetai_def} we have that $\Theta_i+\Theta_j=\Delta_k$ and hence the condition $\Theta_i+\Theta_j>\etah$ is nothing else than the convergence condition for the 2pt-function integral.} In that case we find $\Theta_i+\Theta_j-\etah=-\frac{1}{10}$ which does not lead to a divergence.

Finally, let us return to the assumption we made in \eqref{eq:non_overlapping_assumption}. Using the explicit KK spectrum in~\eqref{eq:DeltaKK} we find that for almost all scalar KK modes the condition \eqref{eq:non_overlapping_assumption} is obeyed. The only exceptions are at $p=1$ for  the tower of modes with $b=-2$ and KK numbers $(\ell_1,\ell_2,\ell_1+\ell_2 - 2(2n-1))$ for any positive integer $n$. It will be very interesting to understand whether the supergravity cubic couplings for these KK modes are non-vanishing and, if they are, to revisit the derivation of the 3pt-function~\eqref{eq:3pt_Appell}.

%%%%
\subsection{Consistency checks}
\label{subsec:consistency}
%%%%

We now describe some consistency checks of~\eqref{eq:3pt_Appell} that are based on basic physical requirements. First, we verify that~\eqref{eq:3pt_Appell} obeys the expected permutation symmetry under exchange of operators, which is manifest in the original integral expression~\eqref{eq:3pt_int}. We then consider three physically important limits of the various parameters in~\eqref{eq:3pt_Appell}, and compare them to general expectations.\footnote{We note that for $p=1,4$ we have $\eta=1$ and one can efficiently evaluate the integral in~\eqref{eq:3ptetaintdef} numerically. We have checked that the results of this numerical integration agree with the analytic expression in~\eqref{eq:3pt_Appell}, see Appendix~\ref{sec:Appendix}.}

%%%%%%%%%%
\subsection*{Permutation symmetry}
%%%%%%%%%%

The integral expression~\eqref{eq:3pt_int} for the 3pt-function is manifestly symmetric under exchange of the three operator insertions, but this symmetry is no longer manifest once we perform the $\eta$-integrals and arrive at our result~\eqref{eq:3pt_Appell}. The reason for this is that in~\eqref{eq:3ptetaintdef} we choose to trivialize the integration over $\xi_z$. Verifying that the final expression~\eqref{eq:3pt_Appell} still obeys the permutation symmetry is therefore a non-trivial consistency requirement. To this end we will show that \eqref{eq:3pt_Appell} is invariant under exchanging the operators at $x\leftrightarrow y$ and $y\leftrightarrow z$, which is enough to establish full permutation symmetry.

We start by considering the $y\leftrightarrow z$ exchange, under which $(A,B)\mapsto(B,A)$. This symmetry is then rather straightforward to check due to the following symmetry property of the Appell function,
\begin{align}\label{eq:Appell_trafo1}
	F_4(a,b;c,d;A,B) = F_4(a,b;d,c;B,A)\,,
\end{align}
which is directly visible from its series representation \eqref{eq:Appell}. With this, the $y\leftrightarrow z$ exchange symmetry of \eqref{eq:3pt_Appell} follows.\footnote{In particular, one can check that in \eqref{eq:3pt_Appell} the first and fourth Appell functions, together with their coefficients, are both mapped to themselves, while the second and third are exchanged.}

Establishing the invariance under the $x\leftrightarrow y$ exchange is more involved, since now we have $(A,B)\mapsto(A/B,1/B)$. There is however an identity of the Appell $F_4$ function which involves precisely such a rational change of variables:\footnote{See e.g. \href{http://dlmf.nist.gov/16.16.E10}{(16.16.10) on DLMF} or \S5.11 in \cite{Erdelyi:1953:HTF1}.}
\begin{equation}\label{eq:Appell_trafo2}
\begin{split}
    F_4(a,b; c,d; e,f) &= \frac{\Gamma(d)\Gamma(b-a)}{\Gamma(d-a)\Gamma(b)}(-f)^{-a} F_4\left(a, a-d+1; c, a-b+1; \frac{e}{f}, \frac{1}{f}\right) \\[3pt]
%%%
    &\quad +\frac{\Gamma(d)\Gamma(a-b)}{\Gamma(d-b)\Gamma(a)}(-f)^{-b} F_4\left(b, b-d+1; c, b-a+1; \frac{e}{f}, \frac{1}{f}\right),
\end{split}
\end{equation}
where we note that the second line is related to the first one by exchanging $a$ and $b$. Using the above with $e=A/B$ and $f=1/B$ allows one to trade an Appell $F_4$ with argument $(A/B,1/B)$ for a sum of two $F_4$'s with standard argument $(A,B)$, resulting in an expression involving a total of eight Appell functions.
Note that also the prefactor $(\rd_{yz}^2)^{\eta-\Theta}$ in~\eqref{eq:3pt_Appell} transforms non-trivially under the $x\leftrightarrow y$ exchange, such that all terms get reshuffled. Once the dust settles, one finds that the eight terms can be grouped into four pairs which have the same Appell function and the same powers of $A$ and $B$ in common. Lastly, by using various gamma function identities one can show equality with~\eqref{eq:3pt_Appell}, thus establishing the invariance under the $x\leftrightarrow y$ exchange.

Altogether, this shows that the 3pt-function result~\eqref{eq:3pt_Appell} is fully symmetric under the exchange of operators, even though this is not manifest from its functional form. This constitutes a first non-trivial consistency check.

%%%%%%%%%%
\subsection*{$\eta\rightarrow0$ limit}
%%%%%%%%%%
This limit corresponds to `removing' the $\mathbb{R}^{\eta}$ subspace of $\R^D$, meaning there is no integration to perform in~\eqref{eq:3pt_int}, and hence we should find that $\langle\O_1(x)\O_2(y)\O_3(z)\rangle_\eta$ reduces to the standard form of a CFT 3pt-function. Indeed, taking the $\eta\rightarrow0$ limit of \eqref{eq:3pt_Appell} yields
\begin{align}
	\langle\O_1(x)\O_2(y)\O_3(z)\rangle_\eta ~\xrightarrow{~\eta\,\rightarrow\,0~}~\frac{\mathcal{V}_{\eta=0}}{(\rd_{xy}^2)^{\Theta_3}(\rd_{xz}^2)^{\Theta_2}(\rd_{yz}^2)^{\Theta_1}}\,,
\end{align}
which is the expected result provided that we regularize the infinite volume factor as $\mathcal{V}_{\eta=0}=1$. Note that when taking this limit, it turns out that only the fourth term in~\eqref{eq:3pt_Appell} contributes: the prefactor $\Gamma(\etah)$ of the fourth Appell function cancels against the same factor in the overall normalization $\N_{\Delta_i}$, c.f.~\eqref{eq:normfactor}. Taking the $\eta\rightarrow0$ limit then boils down to the fact that
\begin{align}\label{eq:limit_F4}
\lim_{b\to0}F_4(a,b;c,d;e,f)=1\,,
\end{align}
which follows from the series representation~\eqref{eq:Appell}.

%%%%%%%%%%
\subsection*{Extremal limit}
%%%%%%%%%%
Next, let us consider an extremal arrangement of operator dimensions, for instance $\Delta_1+\Delta_2=\Delta_3$. This corresponds to the limit $\Theta_3\to0$, while $\Theta_1\to\Delta_2$ and $\Theta_2\to\Delta_1$. The 3pt-function \eqref{eq:3pt_Appell} in this limit then reduces to
\begin{align}\label{eq:3pt_extremal}
	\langle\O_1(x)\O_2(y)\O_3(z)\rangle_\eta ~\xrightarrow{~\Theta_3\,\rightarrow\,0~}~\frac{1}{\Veta}\,\langle\O_1(x)\O_1(z)\rangle_\eta\,\langle\O_2(y)\O_2(z)\rangle_\eta \,.
\end{align}
This factorization into two 2pt-functions is expected from the integral formula~\eqref{eq:3ptetaintdef} together with the expression for the 2pt-function~\eqref{eq:2ptreseta}. This behavior also mirrors a similar factorization in usual AdS/CFT, after proper regularization of the divergent 3pt-function Witten diagram, see e.g. \cite{DHoker:2000xhf}.

To derive the result~\eqref{eq:3pt_extremal} we can proceed similarly to the $\eta\rightarrow0$ limit, where only one of the four terms from \eqref{eq:3pt_Appell} contributes. In the extremal limit with $\Theta_3\to0$ only the second term contributes, because it comes with a prefactor of $\Gamma(\Theta_3)$ which cancels against the same factor in $\N_{\Delta_i}$. Establishing the limit~\eqref{eq:3pt_extremal} then again boils down to using~\eqref{eq:limit_F4}. Similarly, in the other two extremal limits, i.e. $\Theta_2\rightarrow0$ and $\Theta_1\rightarrow0$, it is respectively the third and fourth term which supports the right factorization property \`a la \eqref{eq:3pt_extremal}.

We stress that here we are only discussing the kinematic structure associated with a 3pt-function with an extremal arrangement of the operator dimensions. Whether such extremal 3pt-functions are actually present in non-conformal D$p$-brane holography is a dynamical question which will be determined by the explicit cubic couplings of the corresponding KK supergravity modes. It will be interesting to study this further.

%%%%%%%%%%
\subsection*{OPE limit}
%%%%%%%%%%
Another limit of physical interest is when two operators approach each other, which in conformal theories gives rise to a convergent operator product expansion. The leading term in such a limit in the auxiliary $D$-dimensional CFT reads
\begin{align}
    \langle\O_1(X)\O_2(Y)\O_3(Z)\rangle ~\xrightarrow{~X\,\rightarrow\,Y~}~\frac{C_{\O_1\O_2\O_3}}{|X-Y|^{\Delta_1+\Delta_2-\Delta_3}|Y-Z|^{2\Delta_3}}\,\Big[1+\ldots\Big]\,,
\end{align}
where $C_{\O_1\O_2\O_3}$ is the OPE coefficient and the ellipsis stands for terms which are less singular in $X-Y$. Keeping only this most singular term, and performing the integration as prescribed by~\eqref{eq:3pt_int} yields
\begin{align}\label{eq:OPE_expected}
 \frac{\Veta\pi^\eta\,\Gamma(\Theta_3-\etah)\Gamma(\Theta_1+\Theta_2-\etah)}{\Gamma(\Theta_1+\Theta_2)\Gamma(\Theta_3)}\cdot\frac{C_{\O_1\O_2\O_3}}{|x-y|^{\Delta_1+\Delta_2-\Delta_3-\eta}|y-z|^{2\Delta_3-\eta}}\,.
\end{align}
Taking the auxiliary CFT structure seriously, this is the expected result for the OPE limit in the $d$-dimensional non-conformal theory. We now explain how this arises as a non-trivial limit directly from the 3pt-function $\langle\O_1(x)\O_2(y)\O_3(z)\rangle_\eta$ as given in~\eqref{eq:3pt_Appell}.

For simplicity, consider choosing the operator insertions such that they are on a line. We can then parametrize the limit $x\to y$ as taking $\epsilon\equiv x-y\to0$. We will show later that the leading term in the OPE is not sensitive to any angular dependence on how this limit is taken. We then have $A=\frac{\epsilon^2}{(y-z)^2}$ and $B=\frac{(\epsilon+y-z)^2}{(y-z)^2}$, which appear in the arguments of the Appell functions in~\eqref{eq:3pt_Appell}. Importantly, note that $A$ approaches zero faster than $B$ goes to one, which leads us to first take the $A\to0$ limit of the $F_4$'s under which they reduce to the standard hypergeometric function: to first order in $\epsilon$ one has
\begin{align}
	\lim_{\epsilon\to0}F_4(a,b;c,d;A,B) = \,_2F_1(a,b;d;B)+\ldots\,.
\end{align}
We then take the $\epsilon\to0$ limit on the hypergeometric function, which corresponds to taking $B\to1$ above, and use
\begin{align}\label{eq:2F1limit}
\begin{split}
	\lim_{\epsilon\to0}\,_2F_1(a,b;d;B) &= \frac{\Gamma(d)\Gamma(d-a-b)}{\Gamma(d-a)\Gamma(d-b)}\,\Big[1+\ldots\Big]\\[3pt]
	&~~+ \Big(\frac{2\epsilon}{z-y}\Big)^{d-a-b}\,\frac{\Gamma(d)\Gamma(a+b-d)}{\Gamma(a)\Gamma(b)}\,\Big[1+\ldots\Big].
\end{split}
\end{align}
For the values of parameters appearing in our analysis, we typically have that $d-a-b<0$ and it naively seems that the most singular term in the $\epsilon\to0$ limit comes from the series on the second line in~\eqref{eq:2F1limit}. However, we find that there is a non-trivial cancellation between the terms coming from the first and second, and third and fourth Appell function in \eqref{eq:3pt_Appell}, such that in fact the series of terms on the first line in \eqref{eq:2F1limit} becomes dominant. Then, assuming that $\Theta_3>\etah$, which based on the explicit spectrum of operator dimensions~\eqref{eq:DeltaKK} is always the case for all $p\geq2$ and most $p=0,1$ modes with sub-extremal arrangement of dimensions, the most singular contributions come from the third and fourth term in~\eqref{eq:3pt_Appell}. Adding up these two contributions we then find
\begin{align}\label{eq:OPE_limit}
\begin{split}
 \langle\O_1(x)\O_2(y)\O_3(z)\rangle_\eta~\xrightarrow{~x\,\rightarrow\,y~}~&\frac{\Veta\pi^\eta}{|x-y|^{\Delta_1+\Delta_2-\Delta_3-\eta}|y-z|^{2\Delta_3-\eta}}\times\\[3pt]
 &\qquad\Big[\frac{\Gamma(\Theta_3-\etah)\Gamma(\Theta_1+\Theta_2-\etah)}{\Gamma(\Theta_1+\Theta_2)\Gamma(\Theta_3)}+\ldots\Big],
\end{split}
\end{align}
which perfectly matches the expectation \eqref{eq:OPE_expected}, with $C_{\O_1\O_2\O_3}=1$ as dictated by~\eqref{eq:3ptDdim} and~\eqref{eq:3pt_int}, derived using the OPE limit in the auxiliary $D$-dimensional CFT. Note that since we only used the leading-order term in the $B\to1$ limit in our derivation, any angular dependence on how the $x\to y$ limit is taken drops out from the leading divergence in~\eqref{eq:OPE_limit}.

As a word of caution, we hasten to add that taking the limit $(A,B)\to(0,1)$ in the Appell hypergeometric functions is somewhat subtle as it approaches the boundary of the domain of convergence. It would be interesting to derive the asymptotic expansion in this double limit more rigorously. This would also allow us to explore the form of the omitted sub-leading terms in~\eqref{eq:OPE_limit}. As a further consistency check of the validity of~\eqref{eq:OPE_limit} we have shown that the numerical evaluation of the 3pt-function discussed in Appendix~\ref{sec:Appendix} agrees with~\eqref{eq:OPE_limit}.

%%%%%%%%%%%%
\section{Discussion}
\label{sec:discussion}
%%%%%%%%%%%%

In this paper we studied 2pt- and 3pt-functions of scalar operators in non-conformal QFTs that exhibit an auxiliary conformal structure. This is motivated by the holographic description of the non-conformal D$p$-branes in string theory, which suggests that this structure exists in the large-$N$ and strong-coupling limit of the dual $(p+1)$-dimensional maximally supersymmetric YM theory. Our results raise a plethora of questions and suggest many directions for future work. We discuss some of them below.

\begin{itemize}

\item To derive the kinematic structure of the 2pt- and 3pt-functions in Section~\ref{sec:intCFT}, we used the ad hoc prescription in \eqref{eq:nptprescr} suggested by the auxiliary AdS$_{p+2+\eta}\times S^{8-p}$ uplift of the 10d supergravity solutions~\eqref{eq:NHsugra}. It is important to derive our results directly from the 10d supergravity solutions by obtaining the bulk/boundary propagators of KK modes and computing the corresponding 3pt Witten diagrams.  The holographic renormalization procedure developed in \cite{Kanitscheider:2008kd,Kanitscheider:2009as} should be helpful in this regard in order to regularize the divergences in the bulk supergravity calculation and to determine the normalization coefficients of the 2pt- and 3pt-functions.

\item An important input for the precise holographic calculation of the 2pt- and 3pt-functions for the non-conformal D$p$-branes is the spectrum of operator dimensions $\Delta_i$ and the cubic supergravity coupling for the KK modes. We presented some known values of the operators dimensions of KK modes in \eqref{eq:DeltaKK} but clearly this KK spectroscopy calculation should be performed more systematically and the resulting spectrum should be organized into supermultiplets as dictated by the maximal supersymmetry of the dual SYM theory. This analysis can be performed either by a direct quadratic expansion of the 10d supergravity action around the backgrounds~\eqref{eq:NHsugra}, as was done for AdS$_5\times S^5$ in~\cite{Kim:1985ez}, or by utilizing Exceptional Field Theory (ExFT) techniques along the lines of~\cite{Malek:2019eaz}. The calculation of the cubic couplings of KK supergravity modes by direct methods, as done for instance in \cite{Arutyunov:1999en} for AdS$_5\times S^5$, is more challenging and perhaps the recently developed ExFT approach of \cite{Duboeuf:2023cth} can be adapted to streamline the calculation.

\item To simplify the calculation of quadratic and cubic supergravity couplings one can restrict to the finite subset of KK modes described by the $(p+2)$-dimensional gauged supergravity obtained as a consistent truncation of Type IIA/B supergravity on $S^{8-p}$. The non-linear actions of these supergravity theories are explicitly known and can be readily expanded around their maximally supersymmetric vacua given by the dimensional reduction of the 10d solution~\eqref{eq:NHsugra}. Performing this analysis will not only result in explicit expressions for some of the correlators in the dual SYM theory, but will be an important ingredient for the potential application of ExFT methods to this problem. A concrete model to implement this analysis is provided by the 4d ${\rm ISO}(7)$ gauged supergravity of \cite{Hull:1984yy}, relevant for the holographic description of D2-branes and their dual 3d SYM theory. As will be discussed in \cite{BBGMP} this 4d supergravity action can be used, in conjunction with the results in Section~\ref{sec:intCFT} above, to derive explicit expressions for the 2pt- and 3pt-functions in the SYM theory.

\item The maximally supersymmetric YM theory has 16 supercharges which should lead to powerful supersymmetric Ward identities that constrain the correlators in the theory. It will be most interesting to combine the constraints of supersymmetry with the kinematic structure uncovered in this work to further elucidate the properties of the correlators in these gauge theories at strong coupling.

\item Given the utility of the auxiliary conformal structure we used in Section~\ref{sec:intCFT}, it is important to understand whether any of it survives beyond the leading two-derivative supergravity approximation. It is expected that the scaling similarity of the 10d classical supergravity action is broken by the higher-derivative corrections dictated by string theory. Nevertheless, there may be some remnants of the scaling similarity and its associated auxiliary conformal structure, similarly to how the ${\rm SL}(2,\mathbb{R})$ invariance of the type IIB supergravity action is broken to ${\rm SL}(2,\mathbb{Z})$ in string theory.

\item Since the kinematical structure of $n$pt-functions with arbitrary spin in CFT$_D$ is completely fixed by conformal symmetry for $n=1,2,3$, one can try to repeat the analysis of Section~\ref{sec:intCFT} in this more general case. The analysis is trivial for $n=1$ since the CFT$_D$ correlators vanish. We suspect that the calculation is tractable for $n=2$ but it will probably be unwieldy for $n=3$. It will be very interesting to explore this question further and confront the results with explicit Witten diagram calculations in the supergravity background~\eqref{eq:NHsugra}. Further generalizations to $n>3$ appear complicated and may be hard to pursue by direct calculation. To this end it is important to understand what is the OPE structure dictated by the 3pt-functions we have computed and what is the analogue of conformal blocks for these scaling-similar non-conformal QFTs. It will be interesting to study also whether the connection between perturbative Feynman diagrams and 3pt-functions we stumbled upon in our calculations is simply a coincidence or is a more general feature applicable to $n>3$. Our focus here was on correlation functions of single-trace scalar BPS operators dual to supergravity KK modes. It is clearly important to extend this to multi-trace operators, non-BPS operators dual to string modes, as well as heavy operators described by D-branes, like the ones recently discussed in \cite{Batra:2025ivy}. It will also be interesting to study correlation functions involving both local and extended operators like Wilson lines.

\item Our results for the 2pt- and 3pt-functions provide a prediction for specific quantum theories like 3d maximally supersymmetric YM (for $p=2$) and the BFSS quantum mechanics for $p=0$. Needless to say, it will be very interesting to test this prediction with field theory or matrix model methods. Supersymmetric localization on $S^{p+1}$ may be helpful in this regard. One can contemplate a direct calculation of  localized BPS operators on $S^{p+1}$ by an analysis analogous to \cite{Gerchkovitz:2016gxx} in 4d gauge theories. Alternatively, one can aim to calculate the path integral in the presence of non-trivial supersymmetric deformations on $S^{p+1}$, like a squashing of the metric or certain mass deformations. In the limit of small deformation parameters this calculation will provide concrete data for the integrated correlators of the QFT on $S^{p+1}$ which could be confronted with our holographic results in the large-$N$ and strong-coupling limit.

\item The motivation for our analysis stems from the explicit form of non-conformal D$p$-brane supergravity solutions and the scaling similarity they exhibit. Given the control that this structure offers for holographic calculations, it is important to explore the landscape of scaling similar backgrounds in string theory and their holographically dual non-conformal QFT. We hope that our results will be useful in pursuing this goal.

\end{itemize}

%%%%%%%%%%%%%%%%%%%%%%
\section*{Acknowledgments}
%%%%%%%%%%%%%%%%%%%%%%

We are grateful to Pieter Bomans, Siyul Lee, Juan Maldacena, Joe Minahan, Fri\dh rik Freyr Gautason, Jorge Santos, Silviu Pufu, Anayeli Ram\'irez, Jesse van Muiden, and Kostya Zarembo for valuable discussions, and in particular to Xiangwen Guan for bringing \cite{Davydychev:1992mt} to our attention. NB would like to thank Pieter Bomans and Fri\dh rik Freyr Gautason for numerous enjoyable and informative conversations about non-conformal D$p$-branes over the past decade. We are supported in part by FWO projects G003523N, G094523N, and G0E2723N, as well as by the Odysseus grant G0F9516N from the FWO. 

%%%%%%%%%%%%%%%%%%%%%%
\appendix
\section{Numerical implementation of the 3pt-function}\label{sec:Appendix}
%%%%%%%%%%%%%%%%%%%%%%

The definition of Appell's $F_4$ hypergeometric function via the double series \eqref{eq:Appell} is only valid in the region $\sqrt{|A|} + \sqrt{|B|} < 1$, which is where the series converges. In position space, recalling the identification \eqref{eq:def_AB}, this condition becomes
\begin{equation}\label{eq:triangle_ineq}
      \rd_{xy} + \rd_{xz} < \rd_{yz} \quad\text{or}\quad |y-x| + |x-z| < |y-z| \,.    
\end{equation}
Generic triples of points $x,y,z \in \R^d$ will obey the triangle inequality and therefore lead to values of $A$ and $B$ that are outside of the domain of convergence. It is clear, then, that an analytic continuation of Appell's $F_4$ function is required in order to make sense of the 3pt-function~\eqref{eq:3pt_Appell}. This analytic continuation and a discussion on the numerical evaluation of~\eqref{eq:3pt_Appell} are the subject of this appendix. We follow~\cite{Alkofer:2009} in our exposition, although the analytic continuation that we discuss was originally found in~\cite{Exton}.\footnote{See also~\cite{Ananthanarayan:2020xut} for a discussion on the analytic continuation of the Appell $F_4$ function.}

%%%%%%%%%%
\subsection*{The physical region}
%%%%%%%%%%

\begin{figure}[h]
    \centering
    \includegraphics[width=0.7\linewidth]{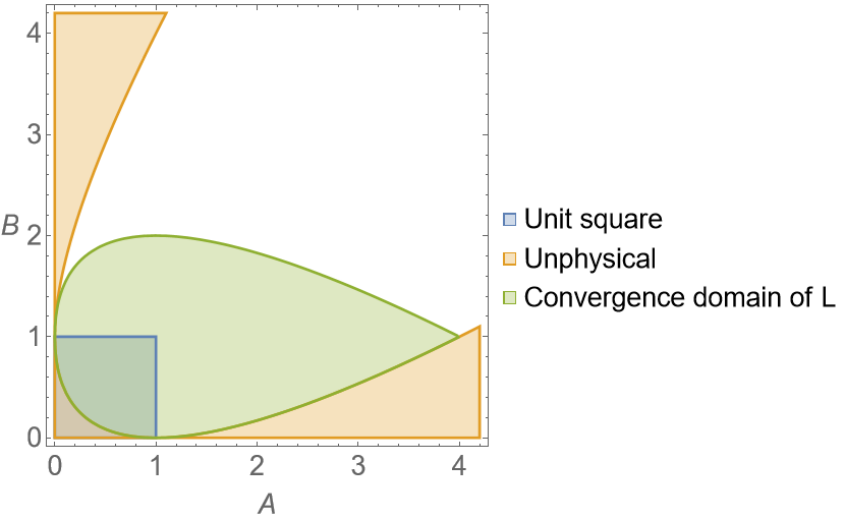}
    \caption{The regions of the $(A,B)$ plane relevant for our analysis.}
    \label{fig:regionsAB}
\end{figure}

We first discuss the region of the $(A,B)$ plane that we are interested in. Since the expression~\eqref{eq:3pt_Appell} is invariant under permutations of $\{x,y,z\}$, we can assume that $y$ and $z$ are the two points that are farthest apart, which, after using~\eqref{eq:def_AB}, implies
\begin{equation}\label{eq:unitsq}
    0 < A < 1\,, \quad 0 < B < 1\,.
\end{equation}
In other words, it suffices to look at the unit square of the $(A,B)$ plane. In this region there are physical and unphysical points. The physical points correspond to distances that obey~\eqref{eq:triangle_ineq} and its permutations. In the unit square, unphysical points are below the curve $\sqrt{A} + \sqrt{B} = 1$, see Figure~\ref{fig:regionsAB}. We thus conclude that the region where we need to find an analytic continuation of the Appell $F_4$ function is
\begin{equation}
    \left\{ (A,B) \mid 0 < A < 1,\; 0 < B < 1,\; \sqrt{A} + \sqrt{B} > 1 \right\}\,.
\end{equation}

%%%%%%%%%%
\subsection*{Analytic continuation of Appell's $F_4$ function}
%%%%%%%%%%

To perform the desired analytic continuation, we start by observing that Appell's $F_4$ function can be expressed as a sum of ordinary hypergeometric functions of one variable:
\begin{equation}\label{eq:expansion_F4_in_2F1s}
    F_4(a,b; c,d; A,B) = \sum_{m=0}^\infty \frac{(a)_{m}(b)_{m}}{(c)_m} \cdot \frac{A^m}{m!}\, \,_2F_1(a+m,b+m; d; B) \,.
\end{equation}
This allows us to resort to the following analytic continuation of $_2F_1$, see \cite[15.3.6]{Abramowitz:1965}:
\begin{equation}\label{eq:analytic_cont_2F1}
\begin{alignedat}{2}
    & \,_2F_1(a,b; c; z) = \frac{\Gamma(c)\Gamma(c-a-b)}{\Gamma(c-a)\Gamma(c-b)}\, \,_2F_1(a,b; a+b-c+1; 1-z) && \\[3pt]
    & \qquad +\frac{\Gamma(c)\Gamma(a+b-c)}{\Gamma(a)\Gamma(b)}\, (1-z)^{c-a-b} \,_2F_1(c-a,c-b; c-a-b && +1; 1-z) \\[3pt]
    & && (|\operatorname{arg}{(1-z)}| < \pi)  \,.
\end{alignedat}
\end{equation}
In our case, $z \equiv B \in (0,1)$ and thus we manage to avoid a branch cut coming from the principal value of the complex logarithm in the $(1-z)^{\#} = \exp(\#\log(1-z))$ factor of the second line. This branch cut can affect the analysis if we attempt to analytically continue beyond the unit square in the $(A,B)$ plane.

The expression~\eqref{eq:analytic_cont_2F1} can then be inserted in (\ref{eq:expansion_F4_in_2F1s}) to get (we refer the interested reader to \cite{Alkofer:2009} for the details):
\begin{equation}\label{eq:analytic_cont_F4}
\begin{split}
    F_4(a,b; c,d; A,B) & = \frac{\Gamma(d)\Gamma(d-a-b)}{\Gamma(d-a)\Gamma(d-b)}\, 
    G(a, b, 1-d+a, 1-d+b; 1-d+a+b, c; A, 1-B) \\[3pt]
    &\quad +L\left(a,b,c,d; \frac{(1-B)^2}{4A}, \frac{A}{4}\right) \,, 
\end{split}
\end{equation}
where the functions $G$ and $L$ are defined as 
\begin{align}
    & G(\alpha,\beta,\gamma,\delta; \lambda,\mu; U,V) \coloneqq\, \sum_{m=0}^\infty \sum_{n=0}^\infty \frac{(\alpha)_{m+n}(\beta)_{m+n}(\gamma)_{m}(\delta)_{m}}{(\lambda)_{2m+n}(\mu)_{m}} \cdot \frac{U^m}{m!}\frac{V^n}{n!} \,, \\[6pt]
%%%%%
    &L(a,b;c,d;U,V):= \nonumber\\[3pt]
        &\quad \frac{\Gamma(d)\Gamma(a+b-d)}{\Gamma(a)\Gamma(b)}\,\Gamma(c)\Gamma(1/2)\, (-16\,V)^{\frac{d}{2}-\frac{a}{2}-\frac{b}{2}}\,\times \nonumber\\[3pt]
    %%%	
            &\qquad\Biggl\lbrace \frac{1}{\Gamma(\frac{a}{2}+\frac{b}{2}-\frac{d}{2}+\frac{1}{2})\Gamma(c-\frac{a}{2}-\frac{b}{2}+\frac{d}{2})}\,\times \nonumber\\[3pt]
                &\qquad\quad K_{1,3}\left(
                \begin{matrix} 
                    \frac{b}{2}-\frac{a}{2}+\frac{d}{2},\frac{a}{2}-\frac{b}{2}+\frac{d}{2},\frac{a}{2}+\frac{b}{2}-\frac{d}{2},\frac{a}{2}+\frac{b}{2}-c-\frac{d}{2}+1\\[3pt]
                    \frac{b}{2}-\frac{a}{2}+\frac{d}{2},\frac{a}{2}-\frac{b}{2}+\frac{d}{2},\frac{1}{2},\frac{1}{2}
                \end{matrix}\,;\,
    		U,V\right)\nonumber\\[3pt]
            &\qquad +\frac{(-4\,V)^{\frac{1}{2}}}{2(d-a-b+1)\Gamma(\frac{a}{2}+\frac{b}{2}-\frac{d}{2})\Gamma(c-\frac{a}{2}-\frac{b}{2}+\frac{d}{2}+\frac{1}{2})}\,\times \nonumber\\[3pt]
        	&\qquad\quad K_{2,4}\left(
                \begin{matrix}
            		\frac{b}{2}-\frac{a}{2}+\frac{d}{2}+\frac{1}{2},\frac{a}{2}-\frac{b}{2}+\frac{d}{2}+\frac{1}{2},\frac{a}{2}+\frac{b}{2}-\frac{d}{2}-\frac{1}{2},\frac{a}{2}+\frac{b}{2}-c-\frac{d}{2}+\frac{1}{2}\\[3pt]
            		\frac{b}{2}-\frac{a}{2}+\frac{d}{2}-\frac{1}{2},\frac{a}{2}-\frac{b}{2}+\frac{d}{2}-\frac{1}{2},\frac{1}{2},\frac{3}{2}
        		\end{matrix}\,;\,
        		U,V\right)
    	\Biggr\rbrace \nonumber\\[3pt]
    %%%	
        &\quad +\frac{\Gamma(d)\Gamma(a+b-d)}{\Gamma(a)\Gamma(b)}\,\Gamma(c)\Gamma(-1/2)\,(16\,UV)^{1/2}(-16\,V)^{\frac{d}{2}-\frac{a}{2}-\frac{b}{2}-\frac{1}{2}}\,\times \nonumber\\[3pt]
            &\qquad \Biggl\lbrace \frac{1}{\Gamma(\frac{a}{2}+\frac{b}{2}-\frac{d}{2})\Gamma(c-\frac{a}{2}-\frac{b}{2}+\frac{d}{2}-\frac{1}{2})}\,\times \nonumber\\[3pt]
        	&\qquad\quad K_{1,3}\left(\begin{matrix}
            		\frac{b}{2}-\frac{a}{2}+\frac{d}{2}+\frac{1}{2},\frac{a}{2}-\frac{b}{2}+\frac{d}{2}+\frac{1}{2},\frac{a}{2}+\frac{b}{2}-\frac{d}{2}+\frac{1}{2},\frac{a}{2}+\frac{b}{2}-c-\frac{d}{2}+\frac{3}{2}\\[3pt]
            		\frac{b}{2}-\frac{a}{2}+\frac{d}{2}+\frac{1}{2},\frac{a}{2}-\frac{b}{2}+\frac{d}{2}+\frac{1}{2},\frac{3}{2},\frac{1}{2}
        		\end{matrix}\,;\,
        		U,V\right)\nonumber\\[3pt]
            &\qquad+\frac{(-4\,V)^{\frac{1}{2}}}{2(d-a-b+1)\Gamma(\frac{a}{2}+\frac{b}{2}-\frac{d}{2}-\frac{1}{2})\Gamma(c-\frac{a}{2}-\frac{b}{2}+\frac{d}{2})}\,\times \nonumber\\[3pt]
                &\qquad\quad K_{2,4}\left(
                \begin{matrix}
                    \frac{b}{2}-\frac{a}{2}+\frac{d}{2}+1,\frac{a}{2}-\frac{b}{2}+\frac{d}{2}+1,\frac{a}{2}+\frac{b}{2}-\frac{d}{2},\frac{a}{2}+\frac{b}{2}-c-\frac{d}{2}+1\\[3pt]
                    \frac{b}{2}-\frac{a}{2}+\frac{d}{2},\frac{a}{2}-\frac{b}{2}+\frac{d}{2},\frac{3}{2},\frac{3}{2}
        		\end{matrix}\,;\,
        		U,V\right)
    	\Biggr\rbrace
\end{align}
with\footnote{The definitions of $K_{1,2}$ and $K_{3,4}$ being used here are slightly different from those in~\cite{Alkofer:2009}.}
\begin{align}
    & K_{1,3}\left(
    \begin{matrix}
    	\alpha,\beta,\gamma,\delta\\ 
    	\alpha,\beta,\lambda,\mu
    \end{matrix}\,;\,
    U,V\right) \coloneqq \nonumber\\[3pt]
        &\quad \sum_{m=0}^\infty \sum_{n=0}^\infty\, \frac{\Gamma(\alpha+m+n)\Gamma(\beta+m+n)}{\Gamma(\alpha+m-n)\Gamma(\beta+m-n)} \cdot \frac{(\gamma)_{m-n}(\delta)_{m-n}}{(\lambda)_{m}(\mu)_{n}} \cdot \frac{U^m}{m!}\frac{V^n}{n!}\,,\\[6pt]
    & K_{2,4}\left(
    \begin{matrix}
    \alpha,\beta,\gamma,\delta\\ 
    \alpha-1,\beta-1,\lambda,\mu
    \end{matrix}\,;\, 
    U,V\right) \coloneqq \nonumber\\[3pt]
        &\quad  4 \sum_{m=0}^\infty \sum_{n=0}^\infty\, \frac{\Gamma(\alpha+m+n)\Gamma(\beta+m+n)}{\Gamma(\alpha-1+m-n)\Gamma(\beta-1+m-n)} \cdot \frac{(\gamma)_{m-n}(\delta)_{m-n}}{(\lambda)_{m}(\mu)_{n}} \cdot \frac{U^m}{m!}\frac{V^n}{n!}\,,
\end{align}
and convergence is guaranteed in the region\footnote{The limiting factor in terms of convergence is the $L$ term in~\eqref{eq:analytic_cont_F4}, since the domain of convergence of the $K$ series that appear in $L$ is contained in that of $G$.}
\begin{equation}
    \sqrt{\frac{(1-B)^2}{4 A}} + \sqrt{\frac{A}{4}} < 1\,,
\end{equation}
which covers the physical region of the unit square (see Figure \ref{fig:regionsAB}).

%%%%%%%%%%
\subsection*{Examples}
%%%%%%%%%%

\begin{figure}[h]
    \centering
    \begin{subfigure}[b]{0.45\textwidth}
        \includegraphics[width=\textwidth]{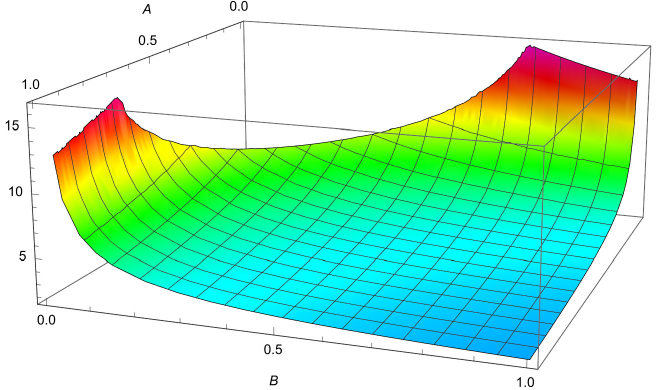}
        \caption{$p=2\,,\, {\rm KK~levels}~\ell_i=(2,2,2)$.}
        \label{fig:p=2_222}
    \end{subfigure}
    \hfill
    \begin{subfigure}[b]{0.45\textwidth}
        \includegraphics[width=\textwidth]{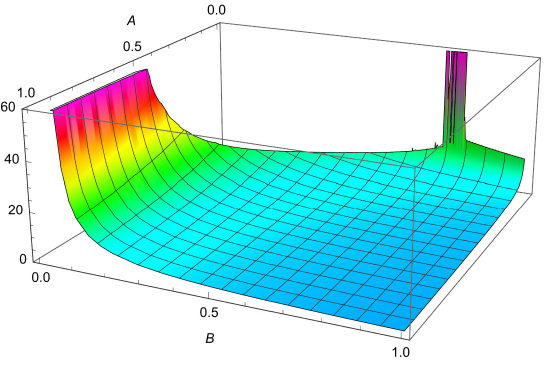}
        \caption{$p=2\,,\, {\rm KK~levels}~\ell_i=(3,3,4)$.}
        \label{fig:p=0_644}
    \end{subfigure}
    \caption{The analytically continued 3pt-function viewed as a function of $A$ and $B$ for $p=2$ and two specific choices of scalar operators.}
    \label{fig:plots_physical_3pt}
\end{figure}

The analytic continuation~\eqref{eq:analytic_cont_F4} described above allows for an efficient numerical evaluation of the 3pt-function~\eqref{eq:3pt_Appell} in the physical region of the $(A,B)$ plane. We have chosen \textit{Mathematica} for this numerical implementation. We focus on 3pt-functions of scalar operators, which have conformal dimensions~\eqref{eq:DeltaKK} with $b=-2$. The scale covariance of the 3pt-function means that we can set $d_{yz} = 1$ in the numerical evaluation. We find that for $p \in \{0,1,2,4\}$ and generic values of the KK levels $\ell_i=(\ell_1, \ell_2, \ell_3)$, the resulting 3pt-function computed using~\eqref{eq:analytic_cont_F4} converges quite fast.\footnote{For some choices of the parameters $(p,\ell_1,\ell_2,\ell_3)$ there are zeroes or poles in the coefficients of the infinite sums used to define the functions $G$ and $L$. In such a situation one has to use different expressions to define the analytic continuation. We have not studied these situations systematically.} In practice, computing the first 20-25 terms of the infinite sums defining the functions $G$ and $L$ is sufficient. 

As an example, Figure \ref{fig:plots_physical_3pt} shows the analytically continued 3pt-function as a function of $A$ and $B$ for two choices of scalar operators for $p=2$, i.e. the 3d SYM theory. The plotted function exhibits singularities at the points $(0,1)$ and $(1,0)$, which correspond respectively to the OPE limits $x \to y$ and $x \to z$, see~\eqref{eq:OPE_limit}. We have studied this singular limit carefully with our numerical implementation and have shown that the numerical results are consistent with the analytic expression in~\eqref{eq:OPE_limit}.

\begin{figure}[h]
    \centering
        \includegraphics[width=0.45\textwidth]{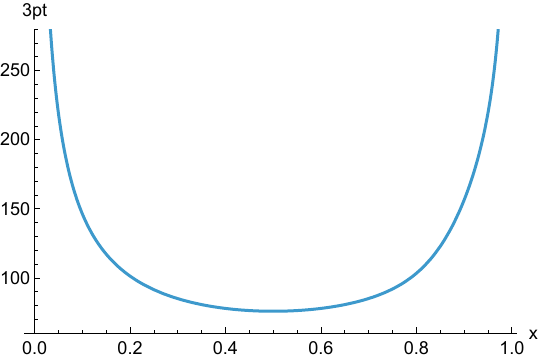}
    \caption{The 3pt-function for $p=0$ and KK levels $\ell_i = (6,4,4)$. Note that without loss of generality we have chosen $y=0$, $z=1$, and $0<x<1$.}
    \label{fig:plot_physical_p0}
\end{figure}

For $p=0$ the space on which the gauge theory is defined is one-dimensional and thus the three operators in the 3pt-function necessarily lie on a line. In this case the values of $A$ and $B$ in the unit square span the curve $\sqrt{A}+\sqrt{B}=1$. For illustration we present the numerical evaluation of a 3pt-function for the D0-brane theory for a specific choice of scalar operators in Figure~\ref{fig:plot_physical_p0}.

\begin{figure}[h]
    \centering
        \includegraphics[width=0.65\textwidth]{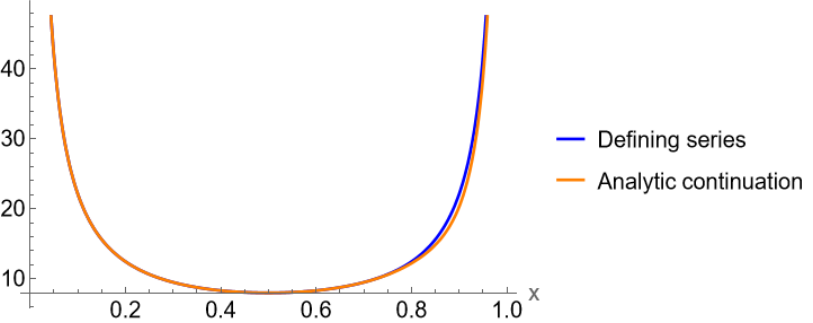}
    \caption{The profile of the 3pt-function~\eqref{eq:3pt_Appell} along the curve $\sqrt{A}+\sqrt{B}=1$ that separates the physical and unphysical regions of the unit square, for the values of the parameters $p=2$ and $\ell_i=(2,2,2)$. The blue curve was evaluated using the defining series of the $F_4$ Appell function \eqref{eq:Appell}, valid in the unphysical region, while for the orange curve we used the analytic continuation discussed in this appendix, valid in the physical one. We have used the parametrization $A=x^2$ and $B = (1-x)^2$.}
    \label{fig:plot_AB1}
\end{figure}

As stressed above, the analytic continuation and numerical implementation is valid in the region~\eqref{eq:unitsq}, while the double sum representation of the Appell $F_4$ function in~\eqref{eq:Appell} is convergent for $\sqrt{A}+\sqrt{B}<1$. Since the curve $\sqrt{A}+\sqrt{B}=1$ is of particular physical interest and defines the boundary of the region of convergence of~\eqref{eq:Appell}, it is worth studying it in some detail. Using a numerical implementation of the double sum in~\eqref{eq:Appell} in the 3pt-function~\eqref{eq:3pt_Appell}, one finds that for $\sqrt{A}+\sqrt{B}=1$ the result agrees very well with the analytically continued expression for the 3pt-function discussed in this appendix. We illustrate this agreement in Figure~\ref{fig:plot_AB1}. We note in passing that in the collinear limit $\sqrt{A}+\sqrt{B}=1$ the $F_4$ Appell function enjoys various identities that relate it to other hypergeometric functions, see e.g. \href{http://dlmf.nist.gov/16.16.E7}{(16.16.7) on DLMF} which applies to our case for $x_{\text{DLMF}}=x$ and $y_{\text{DLMF}}=1-x$ in the parametrization used in Figure \ref{fig:plot_physical_p0}. We have not studied these identities systematically, but they may prove useful for explicit evaluations of the 3pt-function or for a better analytical understanding of the structure of the correlator in the OPE limit.

Lastly, we have performed a simple consistency check on the various formulae above and their numerical implementation. For $p=1,4$ one has that $\eta=1$ and the 3pt-function integral in~\eqref{eq:3ptetaintdef} can be easily evaluated numerically for any three distances $\{\rd_{xy}^2,\rd_{xz}^2,\rd_{yz}^2\}$. We have performed this numerical integration for various values of the dimensions of the operators and a large sample of distances in the unit square region \eqref{eq:unitsq}. We find that the results agree with the numerical evaluation of the 3pt-function \eqref{eq:3pt_Appell} in terms of $F_4$ Appell functions in both the physical and unphysical regions of the unit square.

%%%%%%%%%%%%%%%%%%%%%%%%%%%%
\bibliography{DpCorrelators}
\bibliographystyle{JHEP}
%%%%%%%%%%%%%%%%%%%%%%%%%%%%
\end{document}